\documentclass[journal]{IEEEtran}

\usepackage{cite}
\usepackage[pdftex]{graphicx}
\graphicspath{{figures/}}
\DeclareGraphicsExtensions{.pdf,.jpeg,.png}

\usepackage{amsmath}
\usepackage{algorithmic}
\usepackage{array}
\usepackage{dblfloatfix}

\usepackage{url}

\usepackage{bbm}
\usepackage{algorithm}
\usepackage{xcolor}
\usepackage{caption}
\usepackage{subcaption}
\usepackage{empheq}
\usepackage{amsfonts}

\DeclareMathOperator*{\argmin}{arg\,min}
\DeclareMathOperator*{\argmax}{arg\,max}

\begin{document}

\title{A Bayesian Framework for Digital Twin-Based Control, Monitoring, and Data Collection in Wireless Systems}

\author{
    \IEEEauthorblockN{
        Clement Ruah,\IEEEmembership{~Student Member,~IEEE,}
        Osvaldo Simeone,\IEEEmembership{~Fellow,~IEEE,} and
        Bashir Al-Hashimi\IEEEmembership{,~Fellow,~IEEE}
    }
    \IEEEauthorblockA{
        Department of Engineering, King’s College London, London, UK
    }
    \thanks{
        C. Ruah and O. Simeone are with King’s Communications, Learning \& Information Processing (KCLIP) Lab.
        The work of O. Simeone was supported by the European Research Council (ERC) under the European Union’s Horizon 2020 Research and Innovation Programme (Grant 725732), an Open Fellowship of the EPSRC (reference EP/W024101/1), and the CENTRIC project (Grant 101096379).
        The work of C. Ruah was supported by the Faculty of Natural, Mathematical, and Engineering Sciences at King's College London.
    }
    \thanks{Manuscript received December 1, 2022; revised May 15, 2023.}
}

\IEEEtitleabstractindextext{%
    \begin{abstract}
    Commonly adopted in the manufacturing and aerospace sectors, digital twin (DT) platforms are increasingly seen as a promising paradigm to control, monitor, and analyze software-based, ``open'', communication systems that are expected to dominate 6G deployments. Notably, DT platforms provide a sandbox in which to test artificial intelligence (AI) solutions for communication systems, potentially reducing the need to collect data and test algorithms in the field, i.e., on the physical twin (PT). A key challenge in the deployment of DT systems is to ensure that virtual control optimization, monitoring, and analysis at the DT are safe and reliable, avoiding incorrect decisions caused by ``model exploitation''. To address this challenge, this paper presents a general Bayesian framework with the aim of quantifying and accounting for model uncertainty at the DT that is caused by limitations in the amount and quality of data available at the DT from the PT. In the proposed framework, the DT builds a Bayesian model of the communication system, which is leveraged to enable core DT functionalities such as control via multi-agent reinforcement learning (MARL), monitoring of the PT for anomaly detection, prediction, data-collection optimization, and counterfactual analysis. To exemplify the application of the proposed framework, we specifically investigate a case-study system encompassing multiple sensing devices that report to a common receiver. Experimental results validate the effectiveness of the proposed Bayesian framework as compared to standard frequentist model-based solutions.
    \end{abstract}
    
    \begin{IEEEkeywords}
    Digital Twin, 6G, Reinforcement Learning, Bayesian Learning, Model-based Learning
    \end{IEEEkeywords}
}

\maketitle
\IEEEdisplaynontitleabstractindextext
\IEEEpeerreviewmaketitle

\section{Introduction}

\subsection{Context, Motivation, and Overview} \label{subsec:context_and_motivation}

\begin{figure*}[t]
    \centering
    \includegraphics[width=\textwidth]{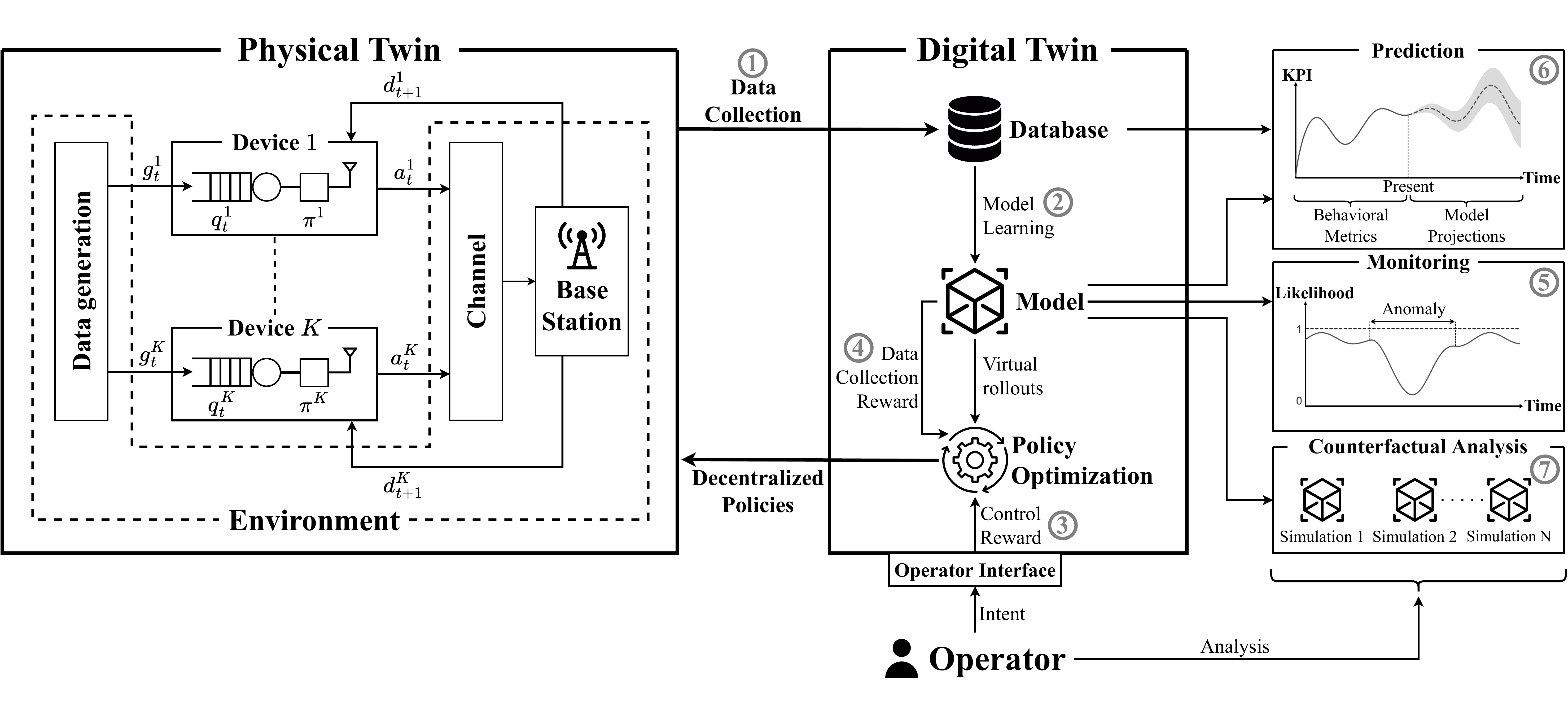}
    \caption{
   A digital twin (DT) platform controlling, monitoring, and analyzing the operation of a communication system operates along the phases of data collection \textcircled{\raisebox{-0.9pt}{1}} (Sec. \ref{subsec:data_collection}), model learning \textcircled{\raisebox{-0.9pt}{2}} (Sec. \ref{sec:model_learning}), policy optimization \textcircled{\raisebox{-0.9pt}{3}} (Sec. \ref{sec:policy_optimization}), and data-collection policy optimization \textcircled{\raisebox{-0.9pt}{4}} (Sec. \ref{subsec:data_collection_optimization}); while also enabling core functionalities such as monitoring \textcircled{\raisebox{-0.9pt}{5}} (Sec. \ref{subsec:anomaly_detection}), prediction \textcircled{\raisebox{-0.9pt}{6}} (Sec. \ref{subsec:prediction}), and counterfactual analysis \textcircled{\raisebox{-0.9pt}{7}} (Sec. \ref{subsec:counterfactual_analysis}).
   Each phase is marked by its corresponding circled number in the figure.
   In the proposed Bayesian framework, the DT maintains a Bayesian model of the communication system, which serves as the physical twin (PT). The Bayesian model quantifies model uncertainty, and enables safe and reliable control, monitoring, and analysis via ensembling and model-disagreement metrics. 
    }
    \label{fig:DT_workflow_MAC}
\end{figure*}

A digital twin (DT) platform is a cyberphysical system in which a physical entity, referred to as the physical twin (PT), and a virtual model, known as the DT, interact based on an automatized bi-directional flow of information \cite{grieves2017digital, kritzinger2018digital}.
Leveraging data received from the PT, the DT maintains an up-to-date model of the PT \cite{glaessgen2012digital},  
which is used to control, monitor, and analyze the operation of the PT \cite{almasan2022network}.
DT platforms are increasingly regarded as an enabling technology for wireless cellular systems built on the open networking principles of disaggregation and virtualization \cite{akman2020oran}, which are expected to be central to 6G \cite{khan2022digital}.
Notably, through the available PT model, DT platforms provide a sandbox in which to test algorithms, protocols, and artificial intelligence (AI) solutions for communication systems, potentially reducing the need to collect data and carry out testing  in the field, i.e., directly on the PT \cite{almasan2022network, hui2022digital}. 

In this regard, a key challenge in the deployment of DT systems is to ensure that virtual control optimization, monitoring, and analysis at the DT are safe and reliable, avoiding incorrect decisions caused by \emph{model exploitation} \cite{sutton2018reinforcement}. To address this challenge, this paper presents a general Bayesian framework with the aim of quantifying and accounting for model uncertainty at the DT that is caused by limitations in the amount and quality of data available at the DT from the PT (see Fig. \ref{fig:DT_workflow_MAC}).

In the proposed framework, the DT builds a \emph{Bayesian model} of the communication system dynamics based on data received from the PT. Unlike conventional \emph{frequentist} parametric models, Bayesian models can quantify model uncertainty by maintaining a distribution over the model parameters \cite{mackay2003information,simeone2022machine}.
This enables \emph{ensembling}-based  control, prediction, and analysis methods, whereby policies, predictions, and recommendations are obtained by accounting for the agreements and disagreements among several models that are consistent with the available information. 
Intuitively, when different models tend to  disagree significantly on an output, this can be taken as quantifiable evidence of model uncertainty.  While ensembling is routinely used in fields such as weather prediction \cite{palmer2022primacy}, its application to DT platforms is still largely unexplored, even outside the field of telecommunications \cite{chen2022advanced, thelen2022comprehensive2}.  

The Bayesian model at the DT can naturally incorporate domain knowledge about the communication systems, including traffic and channel models, while enabling data-driven exploration of the system dynamics. With the available Bayesian model, the DT can carry out the core functionalities of control, monitoring, prediction, data-collection optimization, and counterfactual analysis, while providing uncertainty-aware outputs. We specifically investigate and detail control via model-based Bayesian multi-agent reinforcement learning (MARL), monitoring for anomaly detection, prediction of unobserved dynamics with uncertainty quantification, and data-collection optimization via directed model-based exploration. 

As a possible embodiment of the proposed approach, the DT platform may be implemented as an xApp, or as a collection of connected xApps, that run in the near-real-time RAN Intelligent Controller (RIC) of an Open-RAN (O-RAN) architecture \cite{lacava2022programmable}. As an exemplifying case study, we consider a multi-access PT system consisting of a radio access network (RAN) similar to that studied in \cite{kassab2020multi, valcarce2021toward, miuccio2022learning}.
It is emphasized that, unlike \cite{kassab2020multi, valcarce2021toward, miuccio2022learning}, our goal here is not to address a particular task via MARL, but rather to introduce a general framework supporting the implementation of multiple functionalities at the DT, including control via MARL, monitoring, prediction, and data-collection optimization, despite the limited data transfer from the PT to the DT.

\subsection{Related Work}

This section provides a short review of related papers.

\subsubsection{DT platforms for communication systems}

Position papers advocating for the use of DT platforms for the management of next-generation wireless systems include \cite{khan2022digital, han2022digital, bariah2022interplay, tariq2022toward, luan2021paradigm}.
Specific contributions to the design of DT platforms for wireless systems have investigated mechanisms for DT-PT synchronization \cite{hashash2022edge, hashash2022towards}, DT-aided network optimization and monitoring \cite{hui2022digital}, DT-based control for computation offloading via model-based RL \cite{dai2020deep, zhou2022digital, wang2022digital}, user association \cite{dong2019deep}, as well as the design of intelligent reflecting surfaces \cite{sheen2020digital}.
A layered deployment strategy for DTs from edge to cloud in 5G networks is studied in \cite{jagannath2022digital}; while the
optimization of DT deployment subject to resource and latency constrains in edge servers is investigated in \cite{lu2021adaptive}.
For general reviews on DT systems, we refer the reader to \cite{kritzinger2018digital, thelen2022comprehensive1, thelen2022comprehensive2}.
To the best of our knowledge, the adoption of a general Bayesian framework for the development of DT platforms implementing control, monitoring, and analysis functionalities is yet to be proposed.

\subsubsection{Model-based reinforcement learning}

Reinforcement learning (RL) algorithms fall into two categories: \emph{model-free} algorithms, in which the policy is optimized through trial and error interactions with the ground-truth environment, and \emph{model-based} algorithms, where a model of the environment dynamics is first learned, and then used to optimize the policy in a simulated environment \cite{sutton2018reinforcement}.
In the context of DT platforms, model-based algorithms are the natural choice \cite{dai2020deep, matulis2021robot, park2021digital, wang2022digital}. In fact, they allow the DT to optimize policies to be run at the PT, while bypassing the additional communication overhead and potential safety hazards caused by the interactions with the environment required by model-free methods \cite{cronrath2019enhancing}.
That said, DT-aided control can also benefit from model-free RL, e.g., to refine a policy trained based on an inaccurate model \cite{cronrath2019enhancing}.
Conversely, model-free RL can benefit from the DT model by exploring alternative actions inside the DT simulation in-between training steps in the ground-truth environment \cite{wang2022digital}.
In addition, the learned dynamics also serve other core DT functionalities, such as monitoring, prediction and counterfactual analysis \cite{almasan2022network}.

\subsubsection{MARL for communication systems} \label{subsubsec:marl_literature}
In MARL, each agent is given a partial observation of the global system state \cite{oliehoek2016concise}, and the actions of one agent can influence the state of another, rendering the dynamics non-stationary from the single-agent perspective \cite{laurent2011world}.
Thus, optimizing each agent independently often proves sub-optimal.
State-of-the-art MARL algorithms include \emph{centralized training with decentralized execution} (CTDE) methods \cite{kraemer2016multi}, in which training is done at a central location that optimizes a set of single-agent policies to be deployed at the individual agents. 
CTDE algorithms can be implemented using value-based methods, often relying on value-decomposition networks \cite{sunehag2017value}; using actor-critic methods, typically based on the \emph{centralized critic with decentralized actors} (CCDA) paradigm \cite{lowe2017multi}; or using both methods \cite{wang2020off}.
Identifying which agents contribute to the team's success in cooperative settings is not a trivial task, and is known as the \emph{credit assignment problem}.
To tackle this problem, the COMA algorithm in \cite{foerster2018counterfactual} proposes a counterfactual baseline to reflect how the reward would have changed had the agent taken a different action.
Application of MARL in telecommunications can be found in medium access control signaling protocols \cite{mota2021emergence}, dynamic spectrum access \cite{kassab2020multi} and network routing \cite{tao2001multi}.

\subsubsection{Uncertainty quantification in DT platforms} \label{subsubsec:uncertainty_quantification_literature}

Non-stationary dynamics and limited PT-to-DT communication in real-world scenarios may cause the DT to ``desynchronize'' with the ground-truth dynamics of the PT \cite{hashash2022edge}.
In turn, model errors can result in \emph{model exploitation} during policy optimization, whereby the optimized policy takes advantage of inaccuracies in the DT model and behaves sub-optimally with respect to the ground-truth environment.
Therefore, it is critical that the DT reasons explicitly about its epistemic uncertainty regarding its model of the PT \cite{thelen2022comprehensive2} to avoid over-confident and potentially biased decisions.
To this end, references \cite{li2017dynamic} and \cite{yu2021digital} propose to use Bayesian models at the DT.
Unlike our work, the focus of these references is on monitoring and predicting the health status of the PT components.

\emph{Model-based Bayesian RL}, in which the Bayesian model of the environment dynamics reflects the partial observability of the transition probabilities, was investigated in \cite{duff2002optimal, poupart2006analytic} for single-agent applications. A key advantage of Bayesian models in RL is that ensembling techniques support the implementation of well-informed  active exploration, or data-collection, schemes, which target regimes with high epistemic uncertainty \cite{osband2013more, shyam2019model, zhang2021centralized}.

\subsection{Main Contributions}

The main contributions of this paper are as follows.

\noindent $\bullet$ We introduce a Bayesian DT framework for the control, monitoring, and analysis of a communication system. In the proposed framework, the DT maintains a model of the PT dynamics via a distribution over model parameters, supporting ensembling-based  control, prediction of observed and unobserved dynamics, and counterfactual analysis.
The model at the DT can incorporate domain knowledge about the communication systems (see, e.g., \cite{shlezinger2020model}), including traffic and channel models, and is trained based on data collected from the PT.
Data-collection policies can be optimized over successive rounds based on available data at the DT.

\noindent $\bullet$ We investigate and detail the ensemble-based DT functionalities of control via MARL, monitoring for anomaly detection, prediction with uncertainty quantification, and data-collection optimization via directed model-based exploration. 

\noindent $\bullet$ We present an application of the proposed general framework to a multi-access PT system consisting of a RAN. For this system,  we carry experiments that validate the advantages of the proposed Bayesian framework as compared to conventional frequentist model-based approaches for (\emph{i}) optimal control, using performance metrics such as throughput and buffer overflow; (\emph{ii}) anomaly detection, with performance evaluated via the receiving operating curve; (\emph{iii}) prediction of buffer overflow events under a new control policy, assessed via accuracy and calibration metrics; and (\emph{iv}) data-collection optimization, focusing on benefits in terms of data efficiency. 

This work was partially submitted for conference publication as \cite{ruah2022digital}. The conference version presents a partial description of the framework, including only a brief presentation of tabular model learning and of the DT functionality of  anomaly detection. In contrast, this paper provides full details on the proposed framework, encompassing also neural Bayesian learning, data-collection optimization, prediction, and experimental results for data-collection optimization and prediction.

The rest of the paper is organized as follows.
In Sec. \ref{sec:PT_DT_systems}, we describe the PT system under study and its DT.
Sec. \ref{sec:model_learning} covers model learning at the DT, including both tabular and neural network-based approaches. Sec. \ref{sec:policy_optimization} details policy optimization for control, introducing also a solution to the problem of data-collection optimization.
Sec. \ref{sec:monitoring} addresses the monitoring functionalities of anomaly detection, prediction and counterfactual analysis.
The application of the proposed framework to a multi-access system is provided in Sec. \ref{sec:application}, and Sec. \ref{sec:experiments} presents numerical results.
Sec. \ref{sec:conclusion} concludes the paper.

\section{Physical Twin and Digital Twin Systems} \label{sec:PT_DT_systems}

In this paper, we study a Bayesian methodology for the DT-based optimization and monitoring of a telecommunications network, which constitutes the PT.
In this section, we describe the system under study by first providing a general overview of the interactions between the DT and the PT; then detailing the general assumptions made on the ground-truth dynamic model followed by the PT; and finally explaining the parametric model of the PT assumed by the DT.
The next section will then describe the model learning process at the DT.

\subsection{Overview} \label{subsec:overview}

The system under study encompasses a multi-agent PT, which describes a telecommunications network, and a single DT located in the cloud, for a large PT system, or at the edge, for a local PT system \cite{chukhno2022placement}.
The network elements may be mobile devices and/or central units or distributed units of a 5G system \cite{tse2005fundamentals}.
Note that we focus on the case of a single DT, and leave the important problem of coordination among multiple DTs to future work \cite{wu2021digital, jagannath2022digital}.
The DT collects data from the PT, either periodically or in an adaptive manner, and the data is used to optimize a model of the PT dynamics.
The model learned at the DT is used to operate the PT, as well as to provide monitoring functionalities such as anomaly detection, prediction of the PT future possible states, and counterfactual analysis \cite{almasan2022network}.

As detailed in Sec. \ref{subsec:physical_twin}, the PT system under study consists of multiple network elements, such as mobile devices and infrastructure nodes, which are generically referred to as \emph{agents}.
Without loss of generality, the PT system at a given time can be described as being in a specific \emph{state}.
The state of the system may include, for instance, traffic load conditions at radio units and packet queue lengths at the devices.
Furthermore, the PT state evolves over time according to a ground-truth \emph{transition model} that depends on the agents' actions.

As detailed in Sec. \ref{subsec:digital_twin}, the DT collects data from the PT over dedicated periods of time (phase \textcircled{\raisebox{-0.9pt}{1}} in Fig. \ref{fig:DT_workflow_MAC}).
The time interval between two data collection phases may vary, depending also on the result of diagnostic tests at the DT of current PT behavior, e.g., via anomaly detection (see Sec. \ref{subsec:anomaly_detection}).
Based on the data obtained in each data collection period, the DT constructs a model of the transition dynamics of the PT (phase \textcircled{\raisebox{-0.9pt}{2}} in Fig. \ref{fig:DT_workflow_MAC}).

The model is used by the DT to recommend control policies to the PT (phase \textcircled{\raisebox{-0.9pt}{3}} in Fig. \ref{fig:DT_workflow_MAC}), as well as to carry out monitoring functionalities such as anomaly detection (phase \textcircled{\raisebox{-0.9pt}{5}} in Fig. \ref{fig:DT_workflow_MAC}), prediction (phase \textcircled{\raisebox{-0.9pt}{6}} in Fig. \ref{fig:DT_workflow_MAC}), and counterfactual analysis (phase \textcircled{\raisebox{-0.9pt}{7}} in Fig.  \ref{fig:DT_workflow_MAC}).
For example, the control policy may dictate channel access strategies or scheduling algorithms.
We refer to Sec. \ref{sec:application} for a specific instantiation of the framework for a multi-access system.

An essential aspect of the model learned at the DT is the quality of its \emph{uncertainty quantification} \cite{thelen2022comprehensive2}.
In fact, it is critically important for the DT to know what it knows, i.e., to be aware of which operating regimes of the PT are well described by the DT model; as well as to know what it does not know, i.e., to be aware of the operating regimes in which the DT model may fail to correctly describe the operation of the PT.
A \emph{poorly calibrated} DT model, i.e., a model that cannot properly quantify its epistemic uncertainty, may yield unsafe control decisions for the PT; provide incorrect predictions; and fail to recognize abnormal PT behavior \cite{thelen2022comprehensive2}.

Data collection phases in successive periods may be carried out by the PT with the supervision of the DT, which may recommend specific data collection strategies (phase \textcircled{\raisebox{-0.9pt}{4}} in Fig. \ref{fig:DT_workflow_MAC}).
Uncertainty awareness at the DT is also essential for the optimization of the data-collection policy.
In fact, a well-calibrated model enables the DT to assess which operating regimes of the PT call for additional information to be collected to refine or correct the model.

\subsection{Physical Twin} \label{subsec:physical_twin}

The PT system of interest consists of $K$ agents, indexed by integer $k \in \mathcal{K}=\{1, \dots, K\}$, that operate over a discrete time index $t = 1, 2, \dots$
The time index runs over the relevant time units for the system of interest, which are typically time slots or frames.
The agents make decisions at each time $t$ that affect the evolution of the overall state of the system.

Formally, at each time $t$, each agent $k$ takes an \emph{action} $a^k_t$ from a discrete set of possible actions.
For instance, a mobile device may decide whether to transmit or not in a given time slot $t$.
The action is selected by following a \emph{policy} that leverages information collected by the agent regarding the current \emph{state} $s_t$ of the overall system.

The state $s_t$ is a vector encompassing all the variables necessary to describe the evolution of the system from time $t$ onwards.
State variables may be specific to different local parts of the network, and may be functionally and semantically distinct.
For example, a state variable may describe the current traffic conditions at a base station or the quality of the wireless channel on a particular link.
The state $s_t$ evolves according to some \emph{ground-truth transition probability} $T(s_{t+1} | s_t, a_t)$.
Specifically, the probability distribution of the next state $s_{t+1} \sim T(s_{t+1} | s_t, a_t)$ is modelled as a \emph{Markov decision process} (MDP), and only depends on the current state $s_t$ and joint action $a_t = ( a^1_t, \dots, a^K_t )$ of all agents.

At each time $t$, each agent $k$ observes a function $o^k_t$ of the overall state $s_t$.
This captures the fact that an agent $k$ typically has access only to \emph{local} information about the state of the system, such as the buffer queue length for a device or the traffic load for a base station.
We restrict our framework to the case of \emph{jointly observable} states \cite{oliehoek2016concise}, in which the state $s_t$ can be identified based on the collection of the observations $o^k_t$ of all agents $k \in \mathcal{K}$ at time $t$.
Mathematically, this means the state $s_t$ is assumed to be a function of the collection $o_t = ( o^1_t, \dots, o^K_t )$ of all agents' observations.

It is assumed that agents cannot communicate with each other, and thus the overall information available at agent $k$ at time $t$ amounts to its \emph{action-observation history} $h^k_t = ( o^k_1, a^k_1, o^k_2, \dots, a^k_{t-1}, o^k_t )$.
Accordingly, the behavior of agent $k$ is defined by a policy $\pi^k$ that assigns a probability $\pi^k(a^k_t | h^k_t)$ to each possible action $a^k_t$ based on the available information $h^k_t$.
Note that the presented framework is general enough to subsume the case of a single, possibly composite, PT agent by setting $K=1$.
The more general multi-agent setting under study represents well many telecommunication settings of interest (see Sec. \ref{subsubsec:marl_literature}), and has been studied as a use case for DT platforms in, e.g., \cite{vila2023design}.

\subsection{Digital Twin} \label{subsec:digital_twin}

The DT maintains a \emph{model} of the PT ground-truth dynamics $T(s_{t+1} | s_t, a_t)$.
To this end, the DT assumes a family of parametric models $T_{\theta}(s_{t+1} | s_t, a_t)$ that are determined by a parameter vector $\theta$.
In the model learning phase, the parameter vector $\theta$ is optimized based on data collected from the PT.
As we will detail next, the model class $T_{\theta}(s_{t+1} | s_t, a_t)$ should account for any known structure of the PT.
For instance, the DT may be aware that some of the actions in $a_t$ only affect a subset of the state variables in $s_t$.

In order to account for information available at the DT about the structure of the PT, we partition the state $s_t$ into $M$ distinct subsets $\{ s^i_t \}_{i=1}^{M}$ of state variables, such that each subset $s^i_t$ of state variables is a geographically and/or semantically distinct unit.
For instance, a subset $s^i_t$ may correspond to the queue lengths of a subset of devices connected to the same base station; while a subset $s^j_t$, with $j \neq i$, may describe the channel conditions for all devices connected to a base station.

\begin{figure}[t]
    \centering
    \begin{subfigure}{0.49\textwidth}
        \centering
        \includegraphics[keepaspectratio, width=\textwidth, height=6cm]{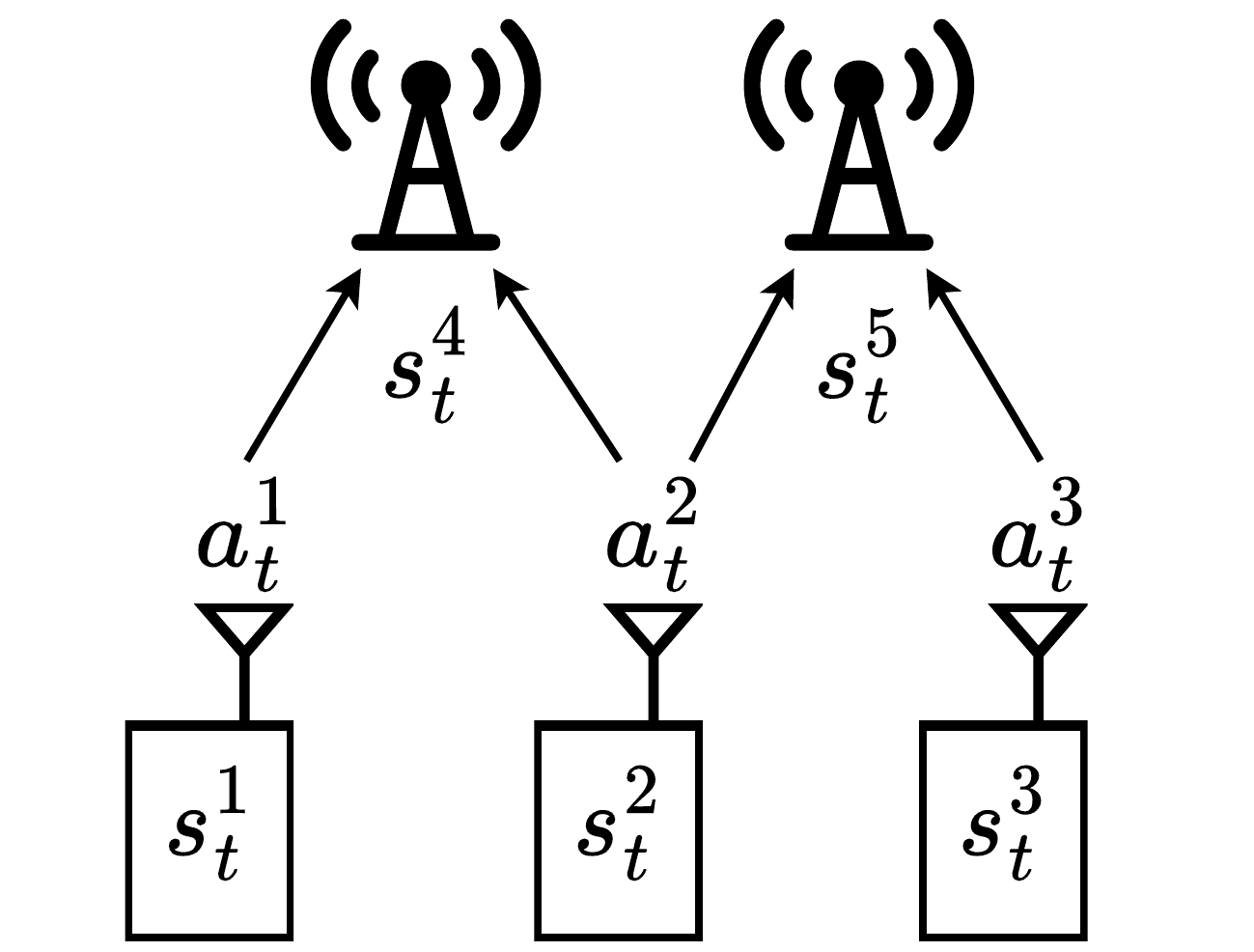}
        \caption{}
        \label{fig:factorization_example_system}
    \end{subfigure}
    \hfill
    \begin{subfigure}{0.49\textwidth}
        \centering
        \vspace{+0.4cm}
        \hspace{-0.5cm}
        \includegraphics[keepaspectratio, width=\textwidth, height=6cm]{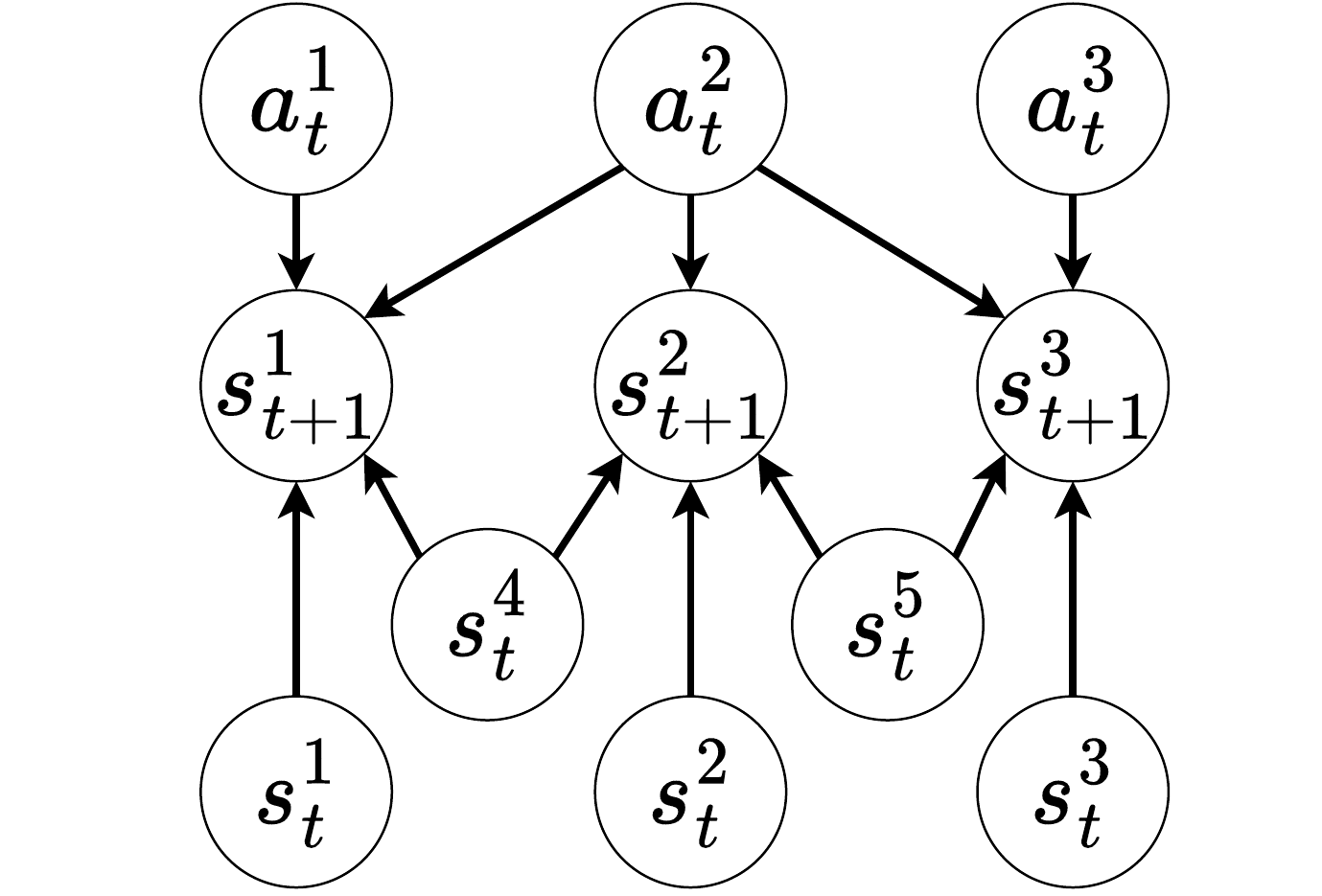}
        \caption{}
        \label{fig:factorization_example_DAG}
    \end{subfigure}
    \caption{
    (a) Example of a PT consisting of three devices and two base stations.
    The internal states $s^1_t$, $s^2_t$ and $s^3_t$ of the three devices may include local battery levels and queue lengths.
    State variables $s^4_t$ and $s^5_t$ describe the propagation conditions on the shared links from the devices to the base stations.
    Note that, in this example, only the second device is in the coverage range of both base stations.
    The actions $a^1_t$, $a^2_t$ and $a^3_t$ of each respective device may include channel access decisions.\\
    (b) Graph representing a possible factorization (\ref{eq:factored_system_dynamics}) assumed at the DT for the state and action variables from time step $t$ to time step $t+1$ for the system described in panel (a).
    Accordingly, the DT assumes that the state of a device at time $t+1$ is affected by the corresponding device's state at time $t$, as well as the actions of the devices connected to the same base station and the channel state for the given base station.
    }
    \label{fig:DAG_example}
\end{figure}

Given the state subset $\{ s^i_t \}_{i=1}^{M}$ and actions $\{ a^k_t \}_{k \in \mathcal{K}}$ of all agents, we introduce a graph with $M$ current state-nodes, one for each subset $s^i_t$; $K$ action-nodes, one for each action $a^k_t$; and $M$ future state-nodes, one for each subset $s^i_{t+1}$.
The graph describes a factorization of the transition probability of the form 
\begin{equation}
\label{eq:factored_system_dynamics}
    T_{\theta}\left( s_{t+1} | s_t, a_t \right) = \prod_{i=1}^{M} T^i_{\theta^i} \left( s^i_{t+1} \middle| s^{>i}_t, a^{>i}_t \right),
\end{equation}
where $s^{>i}_t$ and $a^{>i}_t$ represent the collections of state variables and actions that are considered to directly affect the evolution of state variables in subset $s^i_{t+1}$.
We represent such dependencies by adding a directed edge from action-nodes $a^{>i}_t$ and state-nodes $s^{>i}_t$ to state-node $s^i_{t+1}$.
We refer to Fig. \ref{fig:DAG_example} for an example.
Note that subsets $s^{>i}_t$ and $a^{>i}_t$ may be empty.
For instance, as depicted in the example in Fig. \ref{fig:DAG_example}, state variables that define the channel qualities are generally not affected by the agents' actions (unless solutions such as intelligent reflective surfaces are used \cite{chen2019intelligent}).

We denote as $\theta^i \subseteq \theta$ the subset of model parameters that directly account for the modelled dependence between variables $s^i_{t+1}$ and $(s^{>i}_t, a^{>i}_t)$.
Accordingly, the DT defines $M$ independent parametric models $T^i_{\theta^i} (s^i_{t+1} | s^{>i}_t, a^{>i}_t)$ that, following the factorization in (\ref{eq:factored_system_dynamics}), define the overall dynamic model $T_{\theta}(s_{t+1} | s_t, a_t)$ of the PT with model parameters $\theta = \{ \theta^i \}_{i=1}^{M}$.

A table summarizing the notations can be found in the Appendix \ref{apx:table_notations}.

\section{Model learning at the DT} \label{sec:model_learning}

In this section, we will detail the model learning phase (Fig. \ref{fig:DAG_example} and phase \textcircled{\raisebox{-0.9pt}{2}} in Fig. \ref{fig:DT_workflow_MAC}), during which the DT uses the data collected from the PT to train the model parameters $\theta$ of model (\ref{eq:factored_system_dynamics}).
We first discuss the data-collection phase (phase \textcircled{\raisebox{-0.9pt}{1}} in Fig. \ref{fig:DT_workflow_MAC}), and then present two Bayesian learning methods with different scalability properties.

\subsection{Data Collection} \label{subsec:data_collection}

At the beginning of each data collection phase, the DT may provide the PT with a \emph{data-collection policy} $\pi_d = \{ \pi^k_d(a^k_t | h^k_t) \}_{k \in \mathcal{K}}$, with each agent $k$ receiving policy $\pi^k_d(a^k_t | h^k_t)$.
These policies may be designed by the DT based on information about the PT prior to the data collection phase.
Alternatively, the agents may follow fixed exploration policies, such as distributions $\pi^k_d(a^k_t | h^k_t)$ that assign equal probability to all possible actions $a^k_t$ for each agent.

Starting from an initial state $s_1$ of the PT, all agents in the PT execute the policy $\pi_d$ during $T$ time steps.
After time $T$, each agent $k$ communicates its sequence of observations $\{ o^k_t \}_{t=1}^{T}$ and actions $\{ a^k_t \}_{t=1}^{T}$ to the DT.
Based on this information, and given that the states are assumed to be jointly observable (see Sec. \ref{subsec:physical_twin}), the DT can recover the dataset $\mathcal{D}^{\pi_d}_{T} = \{ (s_t, a_t, s_{t+1}) \}_{t=1}^{T}$ of the $T$ experienced transitions.

Sec. \ref{subsec:data_collection_optimization} will discuss how the DT can optimize the data-collection policy $\pi_d$, while the next subsection covers the model learning phase (phase \textcircled{\raisebox{-0.9pt}{2}} in Fig. \ref{fig:DT_workflow_MAC}).

\subsection{Bayesian Learning} \label{subsec:bayesian_learning}

Based on the dataset $\mathcal{D}^{\pi_d}_{T}$, the DT seeks to optimize the parametric models in (\ref{eq:factored_system_dynamics}) to approximate the ground-truth unknown transition distribution $T(s_{t+1} | s_t, a_t)$.
To this end, we propose that the DT adopts \emph{Bayesian learning} in order to obtain a well-calibrated model.
Bayesian learning aims at evaluating the posterior distribution $P(\theta | \mathcal{D}^{\pi_d}_{T})$ of the unknown model parameters $\theta$.
We define a factorized prior distribution $P(\theta) = \prod_{i=1}^{M} P(\theta^i)$ on the model parameters.
The prior distribution $P(\theta^i)$ can encode both domain knowledge and previous experience obtained from previous data-collection phases.
In particular, in some settings, some of the parameters $\theta^i$ may be known to the DT.
In this case, the prior is concentrated at the known value, and the posterior $P(\theta^i | \mathcal{D}^{\pi_d}_{T})$ trivially coincides with the prior.

Given the factorization in (\ref{eq:factored_system_dynamics}), the posterior distribution $P(\theta | \mathcal{D}^{\pi_d}_{T})$ also factorizes as $P(\theta | \mathcal{D}^{\pi_d}_{T}) = \prod_{i=1}^{M} P(\theta^i | \mathcal{D}^{\pi_d}_{T})$, where the posterior distribution $P(\theta^i | \mathcal{D}^{\pi_d}_{T})$ is given by
\begin{equation}
\label{eq:bayes_formula}
\begin{split}
    P\left( \theta^i \big| \mathcal{D}^{\pi_d}_{T} \right) 
        &\propto P(\theta^i) P\left( \mathcal{D}^{\pi_d}_{T} \big| \theta^i \right) \\
        &= P(\theta^i) \prod_{t = 1}^{T} T^i_{\theta^i}\left( s^i_{t+1} \big| s^{>i}_t, a^{>i}_t \right).
\end{split}
\end{equation}

As we will discuss in the rest of this section, depending on the size of the state and action spaces, computing the exact posterior in (\ref{eq:bayes_formula}) may not be feasible, and one should resort to function approximations.

\subsection{Tabular Bayesian Learning} \label{subsec:tabular_bayesian_learning}

In this subsection, we consider small-scale models, in which:
\noindent $(i)$ the state variable subsets $s^i_t$ take values in a small discrete set $\mathcal{S}^i$;
\noindent and $(ii)$ each conditional distribution $T^i_{\theta^i} (s^i_{t+1} | s^{>i}_t, a^{>i}_t)$ can be expressed as $T^i_{\theta^i} (s^i_{t+1} | x^i_t)$, where $x^i_t$ is a function of variables $(s^{>i}_t, a^{>i}_t)$ that can take a small number of values in a set $\mathcal{X}^i$.
In this case, the parameters $\theta^i$ may be chosen to directly represent the transition probabilities, i.e., we can set $T^i_{\theta^i} (s^i_{t+1} | x^i_t) = \theta^i_{s^i_{t+1} | x^i_t}$ with ${(s^i_{t+1}, x^i_t) \in \mathcal{S}^i \times \mathcal{X}^i}$.
Note that we have the conditions
\begin{equation}
\label{eq:tabular_parameters}
    \sum_{s^i \in \mathcal{S}^i} \theta^i_{s^i | x^i} = 1 \mbox{, and, } \theta^i_{s^i | x^i} \in [0, 1],
\end{equation}
for all $s^i \in \mathcal{S}^i$ and $x^i \in \mathcal{X}^i$.

Exact computation of the posterior distributions $\{ P(\theta^i_{s^i | x^i} | \mathcal{D}^{\pi_d}_{T}) \}_{s^i \in \mathcal{S}^i}$ for $i \in \{1, \dots, M\}$ and $x^i \in \mathcal{X}^i$ can be done using the Dirichlet-Categorical model (see, e.g., \cite{simeone2022machine}).
To this end, we define the prior Dirichlet distribution $P(\{ \theta^i_{s^i | x^i} \}_{s^i \in \mathcal{S}^i}) \sim \mathrm{Dir}(\{ \alpha^i_{s^i | x^i, 0} \}_{s^i \in \mathcal{S}^i})$ with parameters $\alpha^i_{s^i | x^i, 0} > 0$ for $s^i \in \mathcal{S}^i$, such that we have
\begin{equation}
\label{eq:dirichlet_distribution}
    P\left( \left\{ \theta^i_{s^i | x^i} \right\}_{s^i \in \mathcal{S}^i} \right) =
    \frac{
        \prod_{s^i \in \mathcal{S}^i} { \theta^i_{s^i | x^i} }^{\alpha^i_{s^i | x^i, 0} - 1}
    }{
        B\left( \{ \alpha^i_{s^i | x^i, 0} \}_{s^i \in \mathcal{S}^i} \right)
    },
\end{equation}
where the beta function $B(\{ \alpha^i_{s^i | x^i, 0} \}_{s^i \in \mathcal{S}^i})$ is taken as a normalizing constant and depends only on the Dirichlet parameters.
Accordingly, the prior $P(\theta^i)$ factorizes as $P(\theta^i) = \prod_{x^i \in \mathcal{X}^i} P(\{ \theta^i_{s^i | x^i} \}_{s^i \in \mathcal{S}^i})$, and represents prior knowledge or belief about the respective transition model $T^i_{\theta^i}$.
Given the available experience $\mathcal{D}^{\pi_d}_{T}$, the posterior distribution $P(\{ \theta^i_{s^i | x^i}\}_{s^i \in \mathcal{S}^i} | \mathcal{D}^{\pi_d}_{T})$ for $x^i \in \mathcal{X}^i$ is given by the Dirichlet distribution $\mathrm{Dir}(\{ \alpha^i_{s^i | x^i, T} \}_{s^i \in \mathcal{S}^i})$ with the updated parameters
\begin{equation}
\label{eq:dirichlet_update}
    \alpha^i_{s^i | x^i, T} = \alpha^i_{s^i | x^i, 0} + \sum_{t = 1}^{T} \mathbbm{1}_{ \left\{ s^i_{t + 1} = s^i, x^i_t = x^i \right\}}
\end{equation}
for all $s^i \in \mathcal{S}^i$; where the indicator function $\mathbbm{1}_{ \left\{ s^i_{t + 1} = s^i, x^i_t = x^i \right\}}$ is equal to $1$ whenever we have $(x^i, s^i) = (x^i_{t}, s^i_{t+1}) \in \mathcal{D}^{\pi_d}_{T}$, and $0$ otherwise.
Therefore, we update the Dirichlet parameters by counting the number of experienced transitions $(x^i_{t}, s^i_{t+1})$ for all $t \in \{1, \dots, T\}$.

With tabular learning, the number of parameters to be optimized is the same for both frequentist and Bayesian frameworks, with the former relying on maximum likelihood (ML) or maximum a posteriori (MAP) estimates.
This is generally the case in conjugate models, for which the posterior distribution can be evaluated exactly (see, e.g., \cite{barber2012bayesian, simeone2022machine}).

\subsection{Neural Bayesian Learning} \label{subsec:neural_bayesian_learning}

For more complex problems, computing the exact posteriors $P(\theta^i | \mathcal{D}^{\pi_d}_{T})$ in (\ref{eq:bayes_formula}) is typically intractable, and the DT must resort to using approximation methods.
To illustrate this approach, we specifically introduce $M$ neural networks (NNs), one per unknown factor $T^i_{\theta^i}$ in (\ref{eq:factored_system_dynamics}).
For $i \in \{1, \dots, M\}$, the vector $\theta^i$ defines the parameters of the NN $T^i_{\theta^i}$ that takes as input the state and action variables contained in $s^{>i}_t, a^{>i}_t$ at some time $t$, and outputs a probability distribution $T^i_{\theta^i}(s^i_{t+1} | s^{>i}_t, a^{>i}_t)$ over the set of possible states $s^i_{t+1} \in \mathcal{S}^i$ at time $t+1$.

In order to approximate the posterior distribution $P(\theta^i | \mathcal{D}^{\pi_d}_{T})$, we define here a conventional solution based on mean-field variational inference (VI) \cite{kingma2013auto}.
Other approximate inference algorithms, such as Markov chain Monte Carlo (MCMC), would also be applicable \cite{simeone2022machine}.

In the most common implementation of VI, for each factor $i \in \{1, ..., M\}$, one assumes a Gaussian prior given by $P(\theta^i) = \mathcal{N}(\theta^i | 0, \Sigma^i_p)$, where $\Sigma^i_p = \mathrm{Diag}({\sigma^i_{p, 1}}^2, ..., {\sigma^i_{p, P^i}}^2)$ is a diagonal covariance matrix with $\sigma^i_{p, j} > 0$ for $j \in \{1, \dots, P^i\}$, and where $P^i$ denotes the number of parameter in the NN, i.e., the size of vector $\theta^i$.
The posterior $P(\theta^i | \mathcal{D}^{\pi_d}_{T})$ is approximated through the parameterized distribution $q(\theta^i | \phi^i) = \mathcal{N}(\theta^i | \mu^i, \Sigma^i)$ with mean vector $\mu^i = (\mu^i_1, \dots, \mu^i_{P^i})$ and diagonal covariance matrix $\Sigma^i = \mathrm{Diag}({\sigma^i_{1}}^2, ..., {\sigma^i_{P^i}}^2)$, with $\sigma^i_j > 0$ for all $j \in \{1, \dots, P^i\}$.
Variational parameters $\phi^i = (\mu^i_1, \dots, \mu^i_{P^i}, \sigma^i_{1}, \dots, \sigma^i_{P^i})$ are optimized by addressing the problem of minimizing the variational \emph{free energy} \cite{simeone2022machine}, i.e.,
\begin{equation}
\label{eq:min_free_energy_criterion}
\begin{split}
    \argmin_{\phi^i} \mbox{ }
        & \left\{ \mathbbm{E}_{\theta^i \sim q(\theta^i | \phi^i)} \left[ 
            -\log\left( P\left( \mathcal{D}^{\pi_d}_{T} \big| \theta^i \right) \right)
        \right]  \right. \\
        & \left. + \mathrm{KL}\left(
            q(\theta^i | \phi^i) \vert\vert P(\theta^i)
        \right)
        \right\},
\end{split}
\end{equation}
where
\begin{equation}
\label{eq:kl_divergence}
    \mathrm{KL}\left( P(X) \vert\vert Q(X) \right) = \mathbbm{E}_{X \sim P(X)} \left[ \log\left( \frac{P(X)}{Q(X)} \right) \right]
\end{equation}
is the Kullback-Leibler (KL) divergence between two distributions $P$ and $Q$.
The free energy criterion in (\ref{eq:min_free_energy_criterion}) is also known as the negative \emph{evidence lower bound} (ELBO) in the machine learning literature.

While frequentist NNs directly minimize the log-loss $-\log( P(\mathcal{D}^{\pi_d}_{T} | \theta^i) )$ with respect to the $P^i$ parameters in $\theta^i$, optimization of Bayesian NNs via the presented, conventional, VI solution, minimizes the free energy with respect to the $2 P^i$ parameters in vector $\phi^i$.
Using the \emph{reparameterization trick} \cite{kingma2013auto}, problem (\ref{eq:min_free_energy_criterion}) can be addressed iteratively through stochastic gradient descent on an optimization space that encompasses twice the number of parameters as for frequentist learning on the same NN architecture.

\section{Policy Optimization} \label{sec:policy_optimization}

In this section, we discuss the policy optimization phase (phase \textcircled{\raisebox{-0.9pt}{3}} in Fig. \ref{fig:DT_workflow_MAC}), in which the DT leverages the approximate posterior $P(\theta | \mathcal{D}^{\pi_d}_T)$ obtained during the model learning phase (see Sec. \ref{sec:model_learning}) to produce optimal control policies for the multi-agent PT system.
We will also describe the proposed procedure to design efficient data collection policies for the data collection phase (phase \textcircled{\raisebox{-0.9pt}{4}} in Fig. \ref{fig:DT_workflow_MAC}).

\subsection{Setting} \label{subsec:control_policy_optimization}

During policy optimization (phase \textcircled{\raisebox{-0.9pt}{3}} in Fig. \ref{fig:DT_workflow_MAC}), the DT aims at optimizing the \emph{decentralized policy} $\pi = \{ \pi^k(a^k_t | h^k_t) \}_{k \in \mathcal{K}}$ of the $K$ agents so as to maximize some user-specified performance criterion.
This criterion is defined by a \emph{reward} function $r(s_t, a_t, s_{t+1})$, which determines the \emph{total discounted return}
\begin{equation}
\label{eq:total_discounted_return}
    G_t = \sum_{\tau=t}^{+\infty} \gamma^{\tau - t} r(s_\tau, a_\tau, s_{\tau+1}),
\end{equation}
for some exponential discounting factor $\gamma \in [0, 1]$ when the PT applies the policy $\pi$.
The optimal control problem consists of the maximization of the average long-term reward \cite{sutton2018reinforcement}
\begin{equation}
\label{eq:rl_objective}
    \max_{\pi} \mathbbm{E}_{\pi}(G_1).
\end{equation}
This amounts to a Decentralized MDP (Dec-MDP) \cite{oliehoek2016concise}.

We emphasize that the DT has only access to the ensemble of models $T_{\theta}( s_{t+1} | s_t, a_t )$ given by the posterior distribution $P(\theta | \mathcal{D}^{\pi_d}_T)$, and not to the ground-truth distribution $T(s_{t+1} | s_t, a_t)$, when addressing problem (\ref{eq:rl_objective}).
In particular, the DT cannot directly interact with the PT during the policy optimization phase, and must solely rely on the observed data $\mathcal{D}^{\pi_d}_T$.

Given that all policies are issued by the central DT platform, policy optimization can naturally rely on CTDE methods characterized by centralized training at the DT and decentralized execution at the PT.
This class of approaches bypasses non-stationarity issues that affect decentralized learning schemes \cite{laurent2011world}.

\subsection{Control Policy Optimization}

Among possible CTDE methods (see Sec. \ref{subsubsec:marl_literature}), we focus on the COunterfactual Multi-Agent (COMA) algorithm in \cite{foerster2018counterfactual}, a state-of-the-art CCDA method.
The key distinction between the approach adopted here and the conventional COMA implementation is the fact that the model $T_{\theta}( s_{t+1} | s_t, a_t )$ assumed here is stochastic in the sense that the model parameter vector $\theta$ is distributed according to the (approximate) posterior distribution $P(\theta | \mathcal{D}^{\pi_d}_{T})$.

The proposed approach addresses the problem (\ref{eq:rl_objective}) via model-generated virtual rollouts at the DT.
In a manner similar to \cite{zhang2021centralized}, we account for the epistemic uncertainty encoded by the posterior $P(\theta | \mathcal{D}^{\pi_d}_{T})$ by periodically sampling a parameter vector $\theta \sim P(\theta | \mathcal{D}^{\pi_d}_{T})$ during policy optimization so as to produce the next state $s_{t+1} \sim T_{\theta}( s_{t+1} | s_t, a_t )$ in the virtual rollouts.

As is typical in CCDA algorithms \cite{wang2020off}, in a manner similar to standard actor-critic algorithms \cite{sutton2018reinforcement}, the DT maintains a centralized critic $Q_w(s_t, a_t)$, with parameter vector $w$, as well as the decentralized policies $\pi_v = \{ \pi^k_v(a^k_t | h^k_t) \}_{k \in \mathcal{K}}$, with common parameter vector $v$.
During \emph{policy evaluation}, the critic $Q_w(s_t, a_t)$ aims at approximating the \emph{Q-value} $Q^{\pi_v}(s, a) = \mathbbm{E}_{\pi_v}[G_t | s_t=s, a_t=a]$, i.e., the average future return under policy $\pi_v$ starting from a given global state $s$ and joint action $a$.
Then, during \emph{policy improvement}, policies $\pi^k_v(a^k_t | h^k_t)$ for all agents $k \in \mathcal{K}$ are updated to maximize the expected return in (\ref{eq:rl_objective}).
This is done by using the centralized critic $Q_w(s_t, a_t)$ to reward actions that enhance the performance at the system level.
As we will detail next, during the policy optimization phase, we alternate between policy evaluation and policy improvement steps until convergence of the decentralized policy $\pi_v$.
Upon convergence, only the learned policies $\pi^k_v(a^k_t | h^k_t)$ need to be transmitted by the DT to their respective agents.

During policy evaluation, the policy $\pi_v$ is kept constant and the critic $Q_w$ is optimized by leveraging virtual rollouts $(s_1, a_1, r_2, s_2, a_2, ...)$ obtained by following policy $\pi_v$ within model $T_{\theta}( s_{t+1} | s_t, a_t )$.
Since rollouts represent only a finite number of terms in (\ref{eq:total_discounted_return}), the return $G_t$ under policy $\pi_v$ is approximated using the \emph{$n$-step truncated $\lambda$-return} estimator defined as \cite{sutton2018reinforcement}
\begin{equation}
\label{eq:lambda_return_objective}
    G^{\lambda}_{t:t+n} = (1 - \lambda) \sum_{l=1}^{n-1} \lambda^{l-1} G_{t:t+l} + \lambda^{n-1} G_{t:t+n},
\end{equation}
with $\lambda \in [0, 1]$, and
\begin{equation}
\label{eq:l_step_return}
    G_{t:t+l} = \sum_{l'=0}^{l-1} \gamma^{l'} r(s_{t+l'}, a_{t+l'}, s_{t+l'+1}) + \gamma^l Q_{\Bar{w}}(s_{t+l}, a_{t+l}).
\end{equation}
The \emph{target critic} $Q_{\Bar{w}}$ in (\ref{eq:l_step_return}) is used to stabilize the training procedure and shares the same architecture as $Q_w$, with parameters $\Bar{w}$ periodically copied from $w$ \cite{mnih2015human}.
Accordingly, the critic loss function is defined as
\begin{equation}
\label{eq:critic_loss}
    \mathcal{L}_w = \mathbbm{E}_{\pi_v} \left[ \left( G^{\lambda}_{t:t+n} - Q_w(s_t, a_t) \right)^2 \right],
\end{equation}
and the parameters $w$ are obtained iteratively through gradient descent, with target parameters $\Bar{w}$ updated every $N_{\mathrm{target}} \geq 1$ iterations.

After $N_{\mathrm{critic}} \geq 1$ policy evaluation steps, a policy improvement step is carried through gradient ascent with respect to parameters $v$ using the policy gradient theorem with a baseline \cite{sutton2018reinforcement}.
Accordingly, for each agent $k \in \mathcal{K}$, the gradient is given by:
\begin{equation}
\label{eq:policy_gradient}
    \nabla_v J = \mathbbm{E}_{\pi_v} \left[ \sum_{k \in \mathcal{K}} \nabla_v \log\left( \pi^k_v(a^k_t | h^k_t) \right) A^k(s_t, a_t) \right],
\end{equation}
where $A^k(s_t, a_t)$ is the \emph{counterfactual baseline} used by COMA, and defined as
\begin{equation}
\label{eq:coma_baseline}
\begin{split}
    A^k(s_t, a_t) =& Q_w(s_t, a_t) \\
        & - \sum_{a^k \in \{0, 1\}} \pi^k_v(a^k | h^k_t) Q_w\left( s_t, (a^{-k}_t, a^k) \right),
\end{split}
\end{equation}
where $a^{-k}_t = \{ a^{k'}_t \}_{k' \neq k}$ denotes the actions of all agents except agent $k$ at time step $t$.
By marginalizing the contribution of agent $k$ in the baseline, $A^k(s_t, a_t)$ quantifies the effect the action $a^k_t$ of agent $k$ has on the return as compared to its default behavior $a^k_t \sim \pi^k_v(a^k_t | h^k_t)$.
This in turn helps mitigate the credit assignment problem \cite{foerster2018counterfactual}.

In order to encourage exploration of the (virtual) state-action space during the first policy optimization iterations, we draw inspiration from the SAC algorithm \cite{haarnoja2018soft} and use the alternative reward
\begin{equation}
\label{eq:reward_entropy_bonus}
    r_e(s_t, a_t, s_{t+1}) = r(s_t, a_t, s_{t+1}) - \alpha_e \log\left( \pi_v(a_t | s_t) \right),
\end{equation}
with \emph{temperature} hyperparameter $\alpha_e > 0$.
The alternative reward $r_e$ in (\ref{eq:reward_entropy_bonus}) adds an exploration bonus based on the entropy of the policy $\pi_v$, rewarding policies that are stochastic and with high entropy, which in turn enables undirected exploration of the state-action space.
After a given number of policy improvement steps, we return to the original reward definition $r$ until convergence of the control policy $\pi_v$.

\subsection{Data Collection Optimization} \label{subsec:data_collection_optimization}

As discussed in Sec. \ref{subsec:data_collection}, the data-collection policy $\pi_d(a_t | s_t)$ can be optimized by the DT based on the available data and on the DT's assessment about operating regimes characterized by more significant model uncertainty.
For this purpose, the DT uses the available posterior parameter distribution $P( \theta | \mathcal{D}^{\pi_d}_{T} )$ to identify transitions $(s_t, a_t, s_{t + 1})$ that yield high epistemic uncertainty, i.e., where models $T_{\theta}$ drawn from distribution $P( \theta | \mathcal{D}^{\pi_d}_{T} )$ disagree more significantly \cite{simeone2022machine, houlsby2011bayesian}. The resulting disagreement metric is used to engineer a data collection reward $r_d$.
With such reward function, the data collection policy $\pi_d$ is optimized at the DT by following the approach described in Sec. \ref{subsec:control_policy_optimization} with $r_d$ in lieu of $r$.

The reward function $r_d$ should capture the extent to which the ensemble of models $T_{\theta}(s_{t+1} | s_t, a_t)$ with $\theta \sim P( \theta^i | \mathcal{D}^{\pi_d}_{T} )$ disagree on the prediction of the next state $s_{t+1}$ given the previous-step state-action pair $(s_t, a_t)$ \cite{shyam2019model}.
Among the methods presented in Sec. \ref{subsubsec:uncertainty_quantification_literature}, one way to gauge this disagreement is to use the mutual information $\mathrm{I}(s_{t+1} ; \theta | s_t, a_t)$ evaluated under the posterior $P(\theta | \mathcal{D}^{\pi_d}_{T})$ \cite{shyam2019model}.
Accordingly, the data collection reward is defined as 
\begin{equation}
\label{eq:exploration_reward_max}
\begin{split}
    r_d(s_t, a_t) =& \mathcal{H}\left( 
        \mathbbm{E}_{\theta \sim P(\theta | \mathcal{D}^{\pi_d}_{T})}\left[ T_{\theta}(\cdot| s_t, a_t) \right]
    \right)  \\
    & - \mathbbm{E}_{\theta \sim P(\theta | \mathcal{D}^{\pi_d}_{T})}\left[
        \mathcal{H}\left( T_{\theta}(\cdot | s_t, a_t) \right)
    \right],
\end{split}
\end{equation}
where $\mathcal{H}(P(\cdot))= \mathbbm{E}_{s \sim P(s)} \left[ - \log\left( P(s) \right) \right]$ represents Shannon's entropy for the argument distribution.
Note that the entropy terms in (\ref{eq:exploration_reward_max}) are evaluated with respect to the distribution of the next state $s_{t+1}$.
In (\ref{eq:exploration_reward_max}), the first term measures the uncertainty on the next state $s_{t+1}$ for the ensemble model, while the second term represents the average uncertainty associated with each member model $T_{\theta}(s' | s_t, a_t)$ of the ensemble (see also  \cite{simeone2022machine}).

\section{Monitoring Functionalities} \label{sec:monitoring}

In this section, we discuss three typical functionalities that may be run at the DT in addition to control, namely anomaly detection (phase \textcircled{\raisebox{-0.9pt}{5}} in Fig. \ref{fig:DT_workflow_MAC}), prediction (phase \textcircled{\raisebox{-0.9pt}{6}} in Fig. \ref{fig:DT_workflow_MAC}), and counterfactual analysis (phase \textcircled{\raisebox{-0.9pt}{7}} in Fig. \ref{fig:DT_workflow_MAC}). These functionalities are selected as representatives of tasks that are facilitated by the use of uncertainty-aware Bayesian models. 

\subsection{Anomaly Detection} \label{subsec:anomaly_detection}

Anomaly detection aims at detecting significant changes in the dynamics of the PT.
To formulate this problem, assume that, during the operation of the system following policy optimization (phase \textcircled{\raisebox{-0.9pt}{3}} in Fig. \ref{fig:DT_workflow_MAC}), the DT has access to the information $\mathcal{D}^{\pi}_{T^{\mathrm{M}}} = \{ (s_t, a_t, s_{t+1}) \}_{t=1}^{{T^{\mathrm{M}}}}$ about the state-action sequence experienced by the PT within some \emph{monitoring time window} ${T^{\mathrm{M}}}$ under the optimized policy $\pi$.
The DT tests if the collected data $\mathcal{D}^{\pi}_{T^{\mathrm{M}}}$ is consistent with the data reported by the PT during the most recent model learning phase (phase \textcircled{\raisebox{-0.9pt}{1}} and \textcircled{\raisebox{-0.9pt}{2}} in Fig. \ref{fig:DT_workflow_MAC}), or rather if it provides evidence of changed conditions or anomalous behavior.

While frequentist learning is known to perform poorly for detection of out-of-distribution, or abnormal, samples, Bayesian learning has the key advantage of being capable of quantifying epistemic uncertainty via \emph{disagreement-based test metrics}, a property also used in Sec. \ref{subsec:data_collection_optimization} (see, e.g., \cite{daxberger2019bayesian}).
While in Sec. \ref{subsec:data_collection_optimization} disagreement was evaluated on next-state predictions, here the disagreement is defined in terms of the log-likelihood of the observed data. Accordingly, we define as
\begin{equation}
\label{eq:likelihood}
    LL\left( \mathcal{D}^{\pi}_{T^{\mathrm{M}}} \middle| \theta \right) = \sum_{\tau = 1}^{T^{\mathrm{M}}} \log\left( T_{\theta}( s_{t + 1} | s_t, a_t )\pi(a_t | s_t) \right)
\end{equation}
the log-likelihood of model $\theta$ for the reported experience $\mathcal{D}^{\pi}_{T^{\mathrm{M}}}$, where $\pi(a_t | s_t) = \prod_{k \in \mathcal{K}} \pi^k(a^k_t | h^k_t)$.
We then consider the test metric given by the variance
\begin{equation}
\label{eq:monitoring_metric}
\begin{split}
    \mathbbm{E}_{\theta \sim P(\theta | \mathcal{D}^{\pi_d}_{T})} \Bigl[ \Bigl(
        &LL\left( \mathcal{D}^{\pi}_{T^{\mathrm{M}}} \middle| \theta \right) \\
        &- \mathbbm{E}_{\theta \sim P(\theta | \mathcal{D}^{\pi_d}_{T})} \left[
            LL\left( \mathcal{D}^{\pi}_{T^{\mathrm{M}}} \middle| \theta \right)
        \right]
    \Bigr)^2  \Bigr],
\end{split}
\end{equation}
estimated using samples from distribution $P(\theta | \mathcal{D}^{\pi_d}_{T})$.
A larger variance provides evidence of a large epistemic uncertainty, which is taken to indicate an anomalous observation $\mathcal{D}^{\pi}_{T^{\mathrm{M}}}$ as compared to the model learning conditions.

\subsection{Prediction} \label{subsec:prediction}

One of the key motivations behind the model-based approach adopted by the DT paradigm is the possibility of predicting future states of the PT system by simulating the operation of the system via the model. While frequentist models would generally provide unreliable measures of prediction uncertainty, Bayesian models can not only provide useful point predictions but also well-calibrated error bars. 

To describe the problem, we define a \emph{prediction time lag} $T^{\mathrm{H}}$, corresponding to the number of time steps in the future we wish to predict, and a \emph{target metric} $y_p$, which is a function of future trajectories $\mathcal{D}^{\pi}_{T^{\mathrm{H}}} = \{ (s_t, a_t, s_{t+1}) \}_{t=1}^{T^{\mathrm{H}}}$ within the prediction time window duration $T^{\mathrm{H}}$, starting from a known state $s_1$.
We also assume that the agents follow a known policy $\pi$. As an example, the metric of interest $y_p$ may be the average number of packet losses for a subset of devices connected to the same base station over the next $T^H$ time steps (see Sec. \ref{subsec:experimental_results_prediction}). 

Under these conditions, the DT can roll out the model defined by transitions $T_\theta$ and policy $\pi$ in order to estimate statistics of the target metric  $y_p$. With a Bayesian model, such statistics are further averaged over the posterior distribution $P(\theta | \mathcal{D}^{\pi_d}_{T})$, providing a reliable measure of prediction uncertainty.
Accordingly, prediction using a Bayesian model requires a number of samples that is larger as compared to its frequentist counterpart by a factor given by the number of models sampled from the posterior.

\subsection{Counterfactual Analysis} \label{subsec:counterfactual_analysis}

The predictive methodology described in the previous subsection is also a useful tool for counterfactual analysis of the PT behavior \cite{hui2022digital}.
In such analysis, one wishes to assess the impact that changes in the system, as described by the ground-truth dynamics $T$, would have on some target metrics of interest.
To this end, one could roll out different models $T_\theta$ or policies $\pi$ implementing the given changes of interest, and then evaluate measures such as the average treatment effect \cite{wasserman2004all}.

\section{Application to a Multi-Access System} \label{sec:application}

In order to illustrate the operation and the benefits of the proposed framework for the implementation of a DT platform, in the rest of the paper we focus on a multi-access IoT-like wireless network as the PT system to be controlled and monitored \cite{kassab2020multi, valcarce2021toward,miuccio2022learning}. 
This system is implemented as a numerical simulator, which is available on-line \cite{Ruah_Bayesian_Digital_Twin_2022}.

\subsection{Setting} \label{subsec:case_study_setting}

As illustrated in Fig. \ref{fig:DT_workflow_MAC}, the PT system under study comprises $K$ sensing devices that obtain data with correlated data arrivals both in time \cite{nikaein2013simple} and across devices \cite{kassab2020multi}, and communicate with a common base station (BS) over a channel with an unknown distribution. Time is slotted, and each device may transmit in a slot if its buffer is not empty.

With $t$ denoting the time slot index, and following the notation in Sec. \ref{subsec:physical_twin}, each device $k \in \mathcal{K}$ observes its \emph{local state} $o^k_t = ( q^k_t, g^k_t, d^k_t )$, where $q^k_t \in \{0, 1, \dots, Q^k_{\mathrm{max}}\}$ with $Q^k_{\mathrm{max}} \geq 1$ is the number of packets in the device's buffer;
$g^k_t \in \{0, 1\}$ is a binary variable indicating if a new packet is generated ($g^k_t = 1$) at time $t$ or not ($g^k_t = 0$);
and $d^k_t \in \{0, 1\}$ indicates whether a packet sent at the previous time step $t-1$ from device $k$ was successfully delivered at the BS ($d^k_t = 1$) or not ($d^k_t = 0$).
Satisfying the joint observability assumption (see Sec. II-B), the \emph{overall state of the PT} is fully identified given the joint observations of all devices and is represented by $s_t = o_t = (o^1_t, \dots, o^K_t)$.

\subsubsection{Policies} The \emph{access policy} of device $k$ is given by the distribution $\pi^k(a^k_t | h^k_t)$, where we have $a^k_t = 1$ if the device attempts to transmit the first packet in its buffer, and $a^k_t = 0$ if it stays idle during slot $t$.
Finally, we define the (binary) packet-generation vector as $g_t = (g^1_t, \dots, g^K_t)$, the successful packet-delivery vector as $d_t = ( d^1_t, \dots, d^K_t )$, and the packet-transmission vector as $a_t = ( a^1_t, \dots, a^K_t )$.

\subsubsection{Buffers} Each device $k$ maintains a \emph{first-in first-out buffer} of maximum capacity $Q^k_{\mathrm{max}}$, where the buffer state $q^k_t$ evolves according to the deterministic update $P(q^k_{t+1} | q^k_t, d^k_{t+1}, g^k_{t+1})$ given by
\begin{equation}
\label{eq:buffer_update_rule}
    q^k_{t+1} = \min(Q^k_{\mathrm{max}}, q^k_t + g^k_{t+1} - d^k_{t+1}).
\end{equation}
A device $k$ can transmit a packet only if its buffer is not empty, and action $a^k_t$ is automatically set to take value $a^k_t = 0$ otherwise, resulting in the condition $a^k_t \leq q^k_t$.
If device $k$ generates a new packet when the buffer is full and transmission fails, i.e., if we have the equalities $q^k_t = Q^k_{\mathrm{max}}$, $g^k_{t+1} = 1$, and $d^k_{t+1} = 0$, a \emph{buffer overflow} event occurs at time step $t+1$.
In this case, the oldest packet in the buffer is deleted without being sent, and the newly generated packet at time $t+1$ is added to the buffer as per the update rule in (\ref{eq:buffer_update_rule}).

\subsubsection{Packet generation} The \emph{packet generation mechanism} is modelled as a Markov model $P(g_{t+1}|g_t)$.
To account for spatial correlation, we partition the devices into clusters $\{ \mathcal{C}^i \}_{i=1}^{C}$ with $\mathcal{C}^i \subseteq \mathcal{K}$, $\mathcal{C}^i \cap \mathcal{C}^j = \emptyset$ if $i \neq j$ and $\bigcup_{i=1}^{C} \mathcal{C}^i = \mathcal{K}$, where each cluster  $\mathcal{C}^i$ contains devices with correlated packet arrivals.
Accordingly, the data-generation dynamics factorize without loss of generality as
\begin{equation}
\label{eq:data_generation_distribution}
    P(g_{t+1}|g_t) = \prod_{i=1}^{C} P\left( g^{\mathcal{C}^i}_{t+1} \middle| g^{\mathcal{C}^i}_{t} \right),
\end{equation}
where $g^{\mathcal{C}^i}_{t} = \{ g^k_{t} \}_{k \in \mathcal{C}^i}$ for $i \in \{1, \dots, C\}$.

\subsubsection{Channel} The \emph{shared channel} is described by the input-output distribution $P(d_{t+1} | a_t)$, where packet delivery from agent $k$ can be successful ($d^k_{t+1} = 1$) only if a packet was transmitted ($a^k_t = 1$), i.e., we have $a^k_t \geq d^k_{t+1}$.
For each successfully decoded packet, the BS sends back an acknowledgement (ACK) message to the sending device $k$ over an error-free channel on the control plane.

\begin{figure}[t]
    \centering
    \includegraphics[keepaspectratio, width=8.5cm]{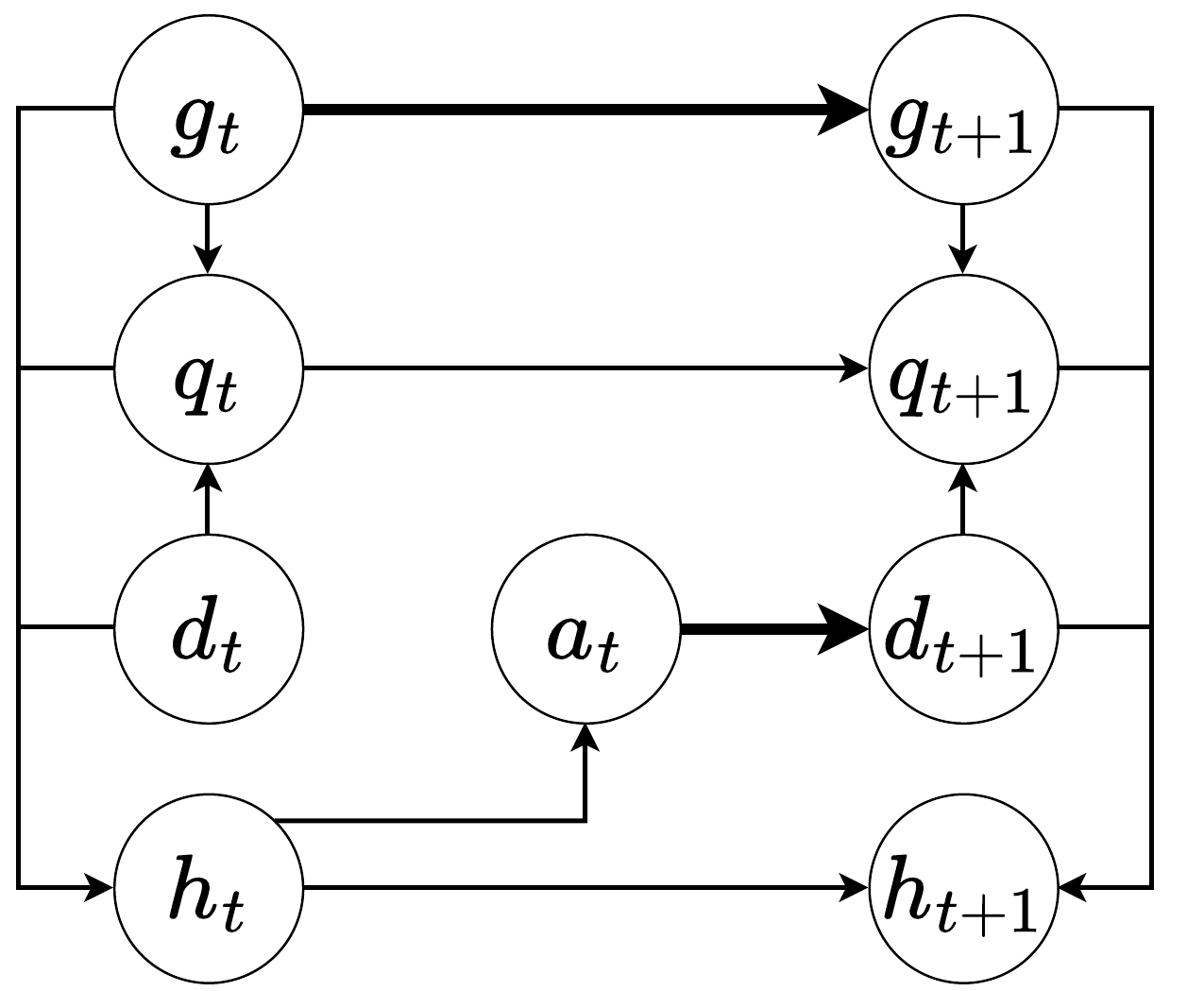}
    \caption{Dependency graph of the multi-access system. Thin lines represent a 1 to 1 relationship per device (independent between devices) while thick lines represent a many to many relationship (correlated between devices)}
    \label{fig:MAC_DBN}
\end{figure}

As an example to be adopted in the next section, in a \emph{multi-packet reception} (MPR) channel, the number of successfully delivered packets $n^{\mathrm{Rx}}_{t+1} = \sum_{k \in \mathcal{K}} d^k_{t+1}$ depends on the number of simultaneous transmissions $n^{\mathrm{Tx}}_t = \sum_{k \in \mathcal{K}} a^k_t$, and the delivered packets are taken uniformly across all the agents that transmit.
Accordingly, the channel distribution is given by \cite{tong2001multipacket}
\begin{equation}
\label{eq:mpr_channel_distribution}
    P(d_{t+1} | a_t) = P(n^{\mathrm{Rx}}_{t+1} | n^{\mathrm{Tx}}_t) \times \frac{\prod_{k \in \mathcal{K}} \mathbbm{1}_{\left\{ a^k_t \geq d^k_{t+1} \right\}}}{\binom{ n^{\mathrm{Tx}}_t }{ n^{\mathrm{Rx}}_{t+1} }}.
\end{equation}

\subsection{DT Model}

Following the system description in the previous section, the DT model assumes the factorization (\ref{eq:factored_system_dynamics}) illustrated in Fig. \ref{fig:MAC_DBN}, which is of the form
\begin{equation}
\label{eq:case_study_dt_factorization}
\begin{split}
T_{\theta}(s_{t+1} | s_t, a_t) = & P_{\theta^{\mathrm{G}}}(g_{t+1}|g_t) \times P_{\theta^{\mathrm{C}}}(d_{t+1} | a_t) \times \\
    & \prod_{k \in \mathcal{K}}  P(q^k_{t+1} | q^k_t, d^k_{t+1}, g^k_{t+1}),
\end{split}
\end{equation}
where the deterministic queue dynamics $ P(q^k_{t+1} | q^k_t, d^k_{t+1}, g^k_{t+1})$ defined by (\ref{eq:buffer_update_rule}) are assumed to be known to the DT, and  the model parameters $\theta = \{\theta^{\mathrm{G}}, \theta^{\mathrm{C}}\}$ determine the packet generation and channel models, respectively.
The DT is also assumed to be aware of the cluster partitions $\{ \mathcal{C}^i \}_{i=1}^{C}$ in (\ref{eq:data_generation_distribution}), e.g., based on the network topology, so that the data generation model $P_{\theta^{\mathrm{G}}}(g_{t+1}|g_t)$ consists of $C$ independent models $T_{\theta^{\mathrm{G}, i}}(g^{\mathcal{C}^i}_{t+1} | g^{\mathcal{C}^i}_t)$ with parameters $\theta^{\mathrm{G}} = \{ \theta^{\mathrm{G}, i} \}_{i=1}^{C}$.
As for the channel, the DT optimizes an MPR model $T_{\theta^{\mathrm{C}}}(n^{\mathrm{Rx}}_{t+1} | n^{\mathrm{Tx}}_t)$ of the unknown ground-truth distribution $P(n^{\mathrm{Rx}}_{t+1} | n^{\mathrm{Tx}}_t)$ of the number of received packets given the number of transmitted packets.

\section{Numerical Results}\label{sec:experiments}

In this section, we present numerical results related to the multi-access system introduced in the previous section.
The main goal is to analyze the advantages of the proposed Bayesian framework at the DT for control, anomaly detection, prediction, and data collection optimization.

\subsection{Setup} \label{subsec:experimental_setup}

Consider $K=4$ sensing devices equipped with a buffer of capacity $Q^k_{\mathrm{max}} = 1$ packet, with all buffers being initially empty. This scenario is of interest for devices that transmit updates, discarding previous packets from the queue as outdated. 
Devices $1$ and $2$ form the cluster $\mathcal{C}^1$, while devices $3$ and $4$ form cluster $\mathcal{C}^2$.
The data generation distribution within each cluster does not depend on previously generated data, and is such that both devices cannot simultaneously generate a packet, with a new packet being generated at either device with probability $0.4$.
This capture a situation in which devices monitor distinct parts of a process, e.g., the location of a target in distinct spatial regions.
The channel allows for the successful transmission of a single packet with probability $1$; while, for two simultaneous transmissions, one packet is received with probability $0.8$ and both packets are received with probability $0.2$.
More than two simultaneous transmissions cause the loss of all packets.

\subsection{Implementation} \label{subsec:experiments_implementation}

\subsubsection{Data Collection}

Unless stated otherwise, we adopt a random data collection policy that sets $\pi_d^k( a^k_t = 1 | h^k_t ) = q_t$ for all $k \in \{1,2,3,4\}$ with probability $q_t$ uniformly and independently selected in the interval $[0, 1]$ at each step $t$.

\subsubsection{Model Learning}

Model learning at the DT is carried using the Categorical-Dirichlet model as described in Sec. \ref{subsec:tabular_bayesian_learning} with all prior Dirichlet parameters set to $0.01$.
The DT adopts a memoryless model $T_{\theta^{\mathrm{G}, i}}(g^{\mathcal{C}^i}_{t+1})$ for the data generation process with model parameters $\theta^{\mathrm{G}, i} = \{ \theta^{\mathrm{G}, i}_{{g'}^{\mathcal{C}^i}} \}_{{g'}^{\mathcal{C}^i} \in \{0, 1\}^{\vert \mathcal{C}^i \vert}}$ for $i \in \{1, 2\}$.
Furthermore, the channel model is defined by the model parameters $\theta^{\mathrm{C}} = \{ \theta^{\mathrm{C}}_{n^{\mathrm{Rx}} | n^{\mathrm{Tx}}} \}_{n^{\mathrm{Rx}} \leq n^{\mathrm{Tx}} \in \{0, \dots, K \}}$.

\subsubsection{Reward}

In a similar manner to \cite{miuccio2022learning}, we assume that the reward in (\ref{eq:total_discounted_return}) takes the form
\begin{equation}
    \label{eq:overall_reward_definition}
    r(s_t, a_t, s_{t+1}) = \sum_{k \in \mathcal{K}} \beta^k r^k(o^k_t, a^k_t, o^k_{t+1}),
\end{equation}
with
\begin{equation}
    \label{eq:agent_reward_definition}
    r^k(o^k_t, a^k_t, o^k_{t+1}) = \left\{
    \renewcommand{\arraystretch}{1}
    \begin{array}{ll}
        + \xi & \mbox{if } d^k_{t+1} = 1 \\
        - \xi & \mbox{if } q^k_t = Q^k_{\mathrm{max}} \mbox{, } g^k_{t+1} = 1 \mbox{ and} \\
            & d^k_{t+1} = 0 \\
        - 1 & \mbox{otherwise},
    \end{array}
    \right.
\end{equation}
where the first condition corresponds to successful packet delivery and the second condition to buffer overflow.
The constants $\{\beta^k\}_{k \in \mathcal{K}}$ and $\xi > 0$ are hyperparameters under the control of the network operator at the DT.
In our experiments, we set $\beta^k = 1$ for all $k \in \{1, 2, 3, 4\}$, $\xi = 50$, and the discount parameter in (\ref{eq:total_discounted_return}) is set to $\gamma = 0.95$.

\subsubsection{Actor and Critic}

The critic $Q_w(s_t, a_t)$ and the policies $\pi^k_v(a^k_t | h^k_t)$ for the COMA algorithm presented in Sec. \ref{subsec:control_policy_optimization} are implemented as feedforward neural networks.
Specifically, the policy $\pi^k_v(a^k_t | h^k_t)$ takes as input its current observation $o^k_t$, along with the positional input $p_t = (t \mod{L})$, where $L=4$ is a hyperparameter, resulting in a policy of the form $\pi^k(a^k_t | o^k_t, p_t)$.
More precisely, each neural network $\pi^k_v(a^k_t | h^k_t)$ outputs $L$ probabilities $\{ \pi^k_v(a^k_t | o^k_t, p) \}_{p=0}^{L-1}$ such that $\pi^k_v(a^k_t | h^k_t) = \prod_{p=0}^{L-1} \mathbbm{1}_{\{p = p_t\}} \pi^k_v(a^k_t | o^k_t, p)$.
Partitioning time into frames of $L$ slots, $\pi^k_v(a^k_t = 1 | o^k_t, p)$ can be interpreted as the probability of sending a packet during slot $p$ within the current frame.
The adoption of more complex policies using recurrent neural networks (RNNs) \cite{zhu2017improving} is left for future work.

\begin{figure}[t]
    \centering
    \begin{subfigure}{0.49\textwidth}
        \centering
        \includegraphics[width=\textwidth]{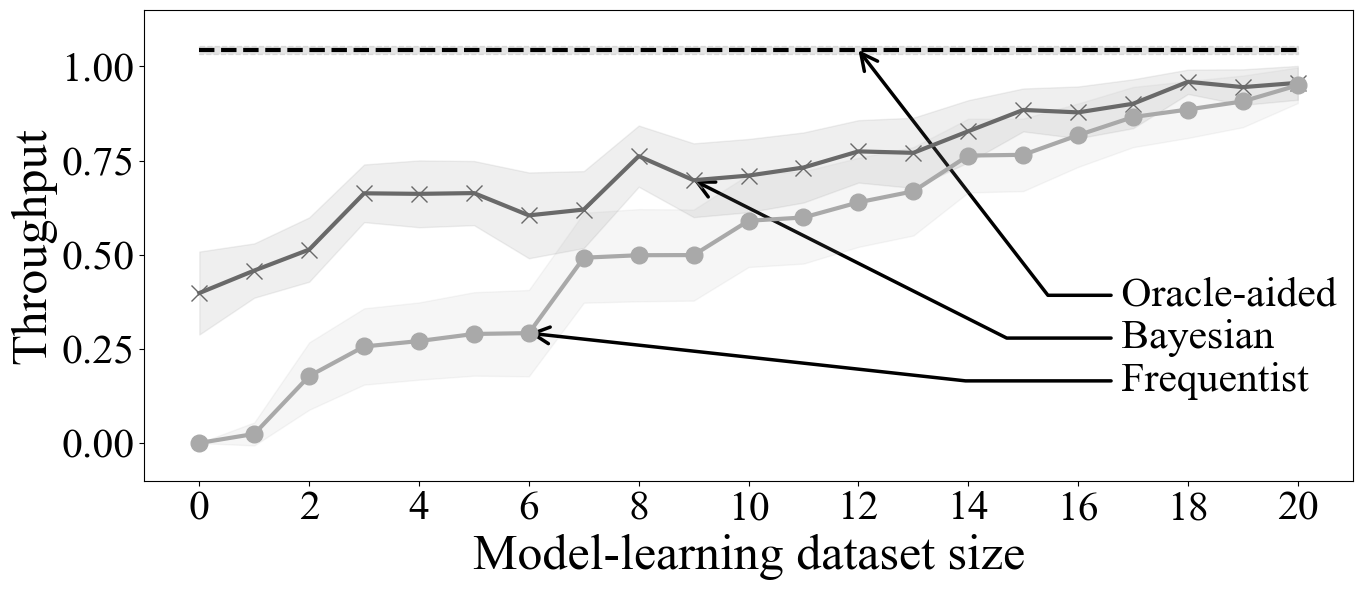}
        \vspace{-0.6cm}
        \caption{}
        \label{fig:throughput_per_model_steps}
    \end{subfigure}
    \hfill
    \begin{subfigure}{0.49\textwidth}
        \centering
        \includegraphics[width=\textwidth]{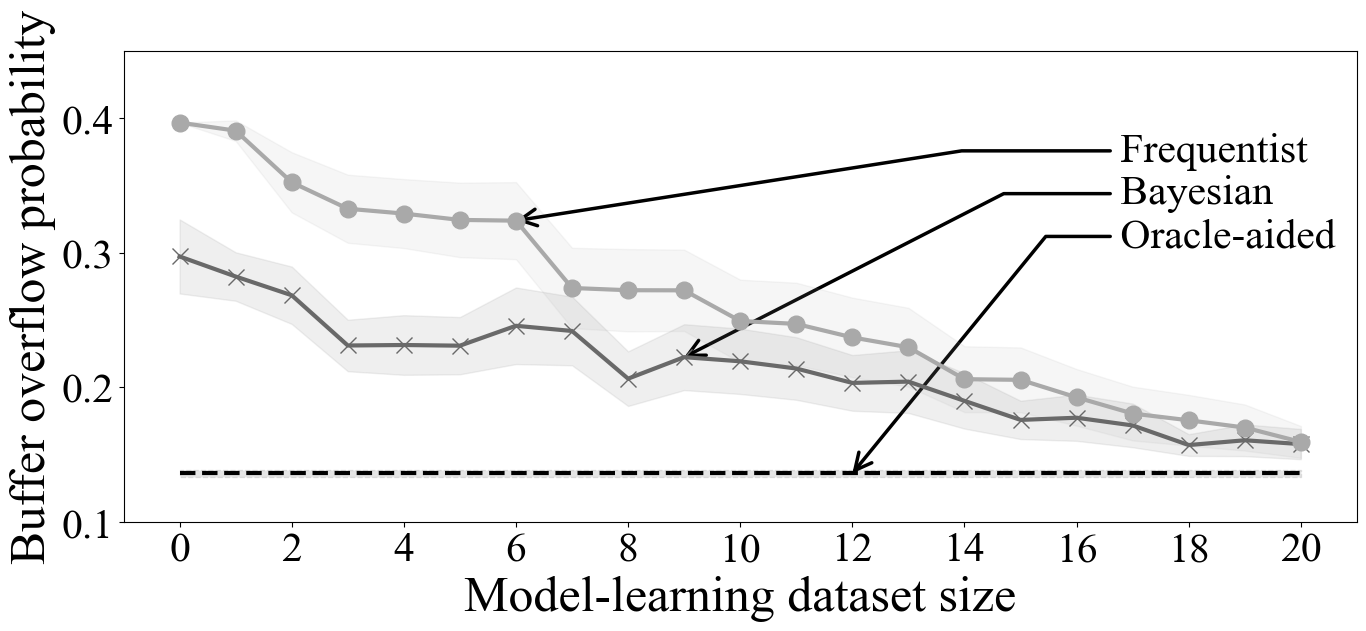}
        \vspace{-0.6cm}
        \caption{}
        \label{fig:overflow_per_model_steps}
    \end{subfigure}
    \caption{
    Throughput (a) and buffer overflow probability (b) as a function of the size of the dataset available in the model learning phase for the proposed Bayesian model-based approach, as well as the oracle-aided model-free and frequentist model-based benchmarks.
    Metrics are averaged over time and over $50$ independent model learning and policy optimization cycles.
    }
    \label{fig:policies_metrics}
\end{figure}

\begin{figure}[t]
    \centering
    \begin{subfigure}{0.49\textwidth}
        \centering
        \hspace{-0.6cm}\includegraphics[keepaspectratio, width=\textwidth, height=7cm]{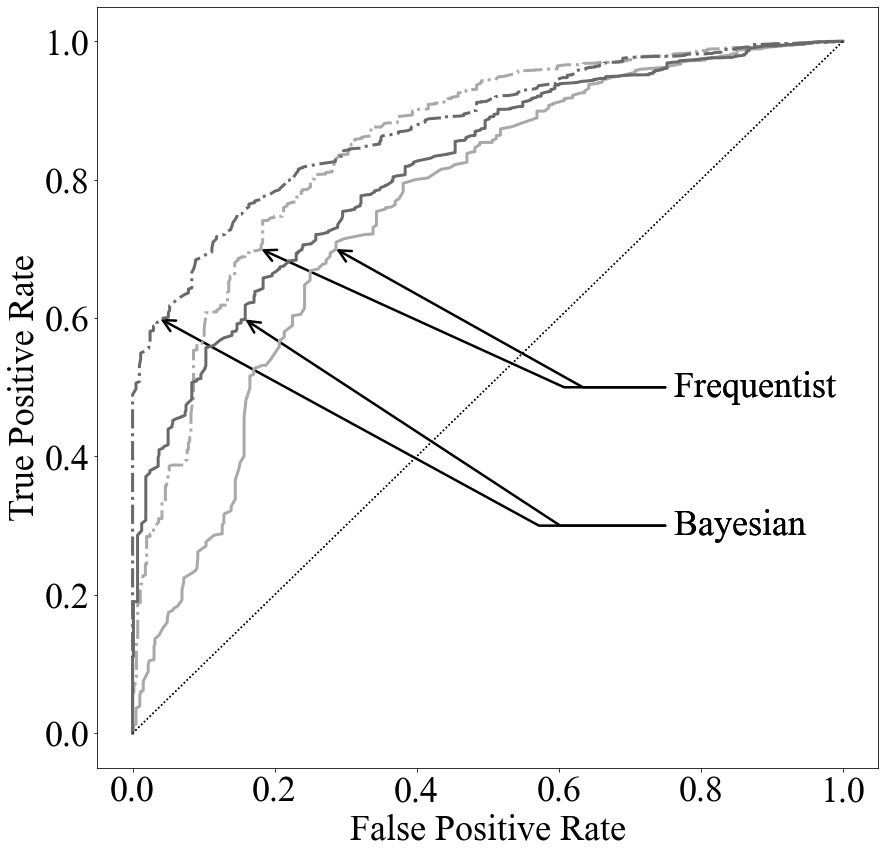}
        \vspace{-0.2cm}
        \caption{}
        \label{fig:roc_anomaly_detection}
    \end{subfigure}
    \hfill
    \vspace{-0cm}
    \begin{subfigure}{0.49\textwidth}
        \centering
        \hspace{-0.6cm}\includegraphics[keepaspectratio, width=0.85\textwidth]{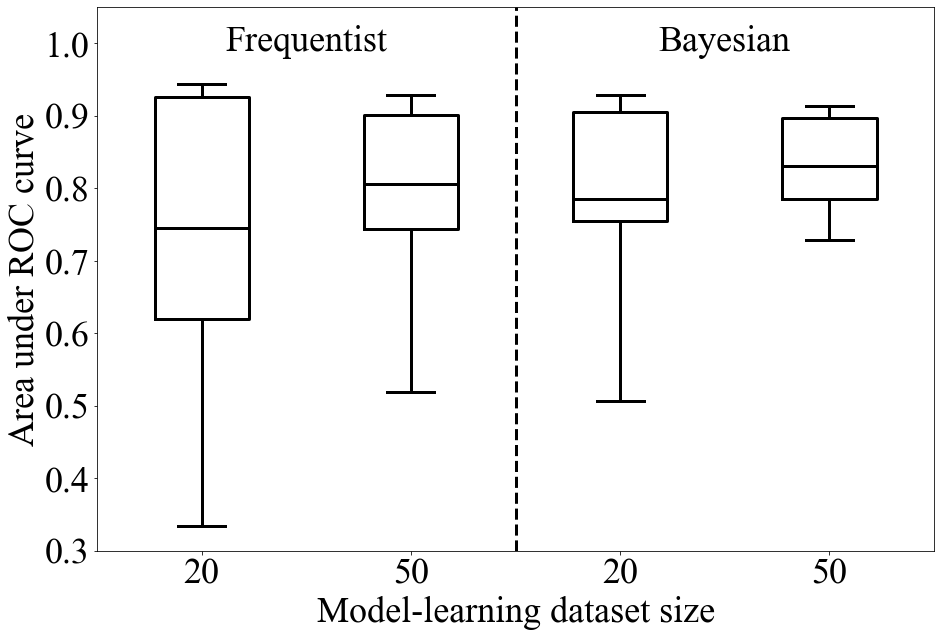}
        \vspace{-0.2cm}
        \caption{}
        \label{fig:auc_anomaly_detection}
    \end{subfigure}
    \caption{
    Mean receiver operating characteristic (ROC) curves (a) and area under ROC curves (AUC) (b) for the Bayesian and frequentist anomaly detection tests.
    Solid lines in (a) represent model learning dataset sizes of $T=20$ steps, while dashed lines correspond to dataset sizes of $T=50$ steps.
    Mean AUCs in (b) are represented by an horizontal bar, while boxes denote the 25\% and 75\% quantiles and whiskers denote the 10\% and 90\% quantiles.
    Results are obtained from $50$ independent data collection and model learning cycles.
    }
    \label{fig:anomaly_detection}
\end{figure}

\subsection{Benchmarks}\label{subsec:experimental_benchmarks}

Throughout the experiments, we compare the performance of the proposed Bayesian framework to the two following benchmarks.
The first is a \emph{frequentist} model-based approach, which obtains a maximum a posteriori (MAP) estimate $\theta^{\mathrm{MAP}} = \argmax_{\theta} P(\theta | \mathcal{D}^{\pi_d}_{T})$ of the model parameter vector $\theta$ during model learning with all Dirichlet prior parameters set to $1.01$. This choice guarantees well-defined solutions for the MAP problem. The frequentist approach uses the single optimized model $T_{\theta^{\mathrm{MAP}}}( s_{t+1} | s_t, a_t )$ for policy optimization, anomaly detection, and prediction.
For policy optimization, we also consider an \emph{oracle-aided} model-free scheme, in which the policy optimizer is allowed to interact with the ground-truth distributions (\ref{eq:data_generation_distribution}) and (\ref{eq:mpr_channel_distribution}) until convergence.

\begin{figure*}[t]
    \centering
    \begin{subfigure}{0.32\textwidth}
        \includegraphics[width=\textwidth]{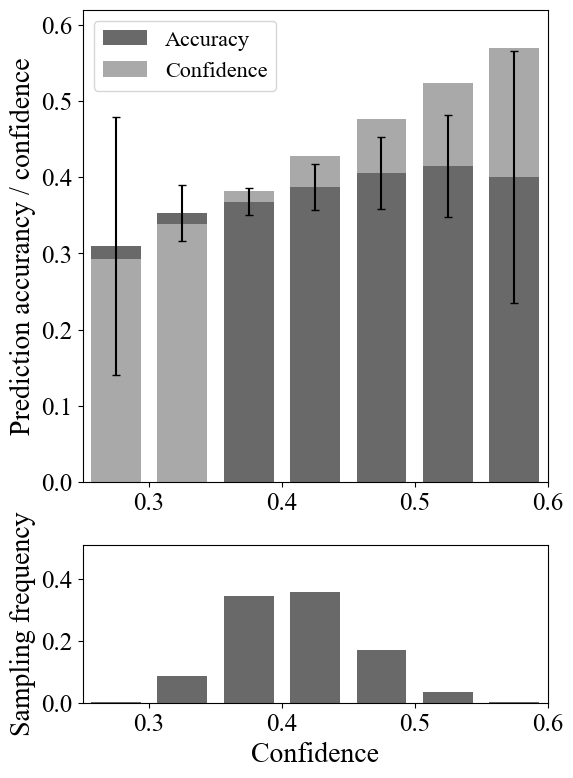}
        \caption{}
        \label{fig:frequentist_reliability_plot}
    \end{subfigure}
    \hfill
    \begin{subfigure}{0.32\textwidth}
        \centering
        \includegraphics[width=\textwidth]{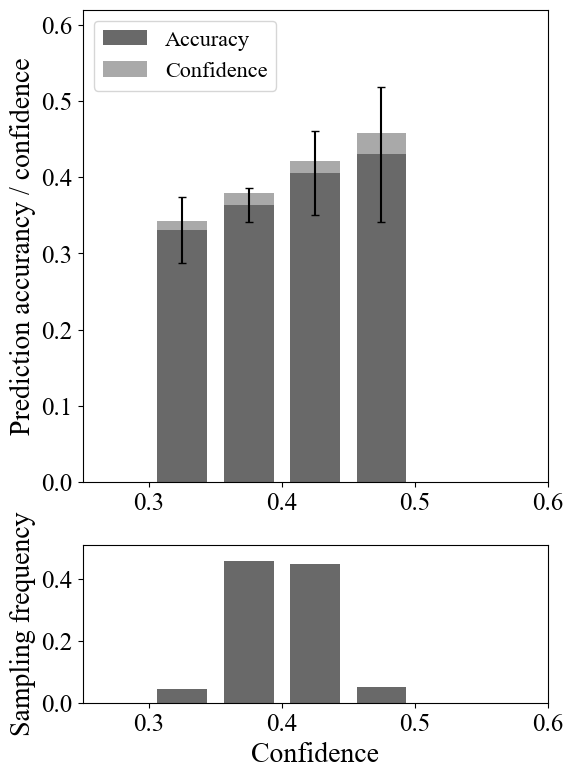}
        \caption{}
        \label{fig:bayesian_reliability_plot}
    \end{subfigure}
    \hfill
    \begin{subfigure}{0.34\textwidth}
        \centering
        \begin{subfigure}{\textwidth}
            \centering
            \includegraphics[width=\textwidth]{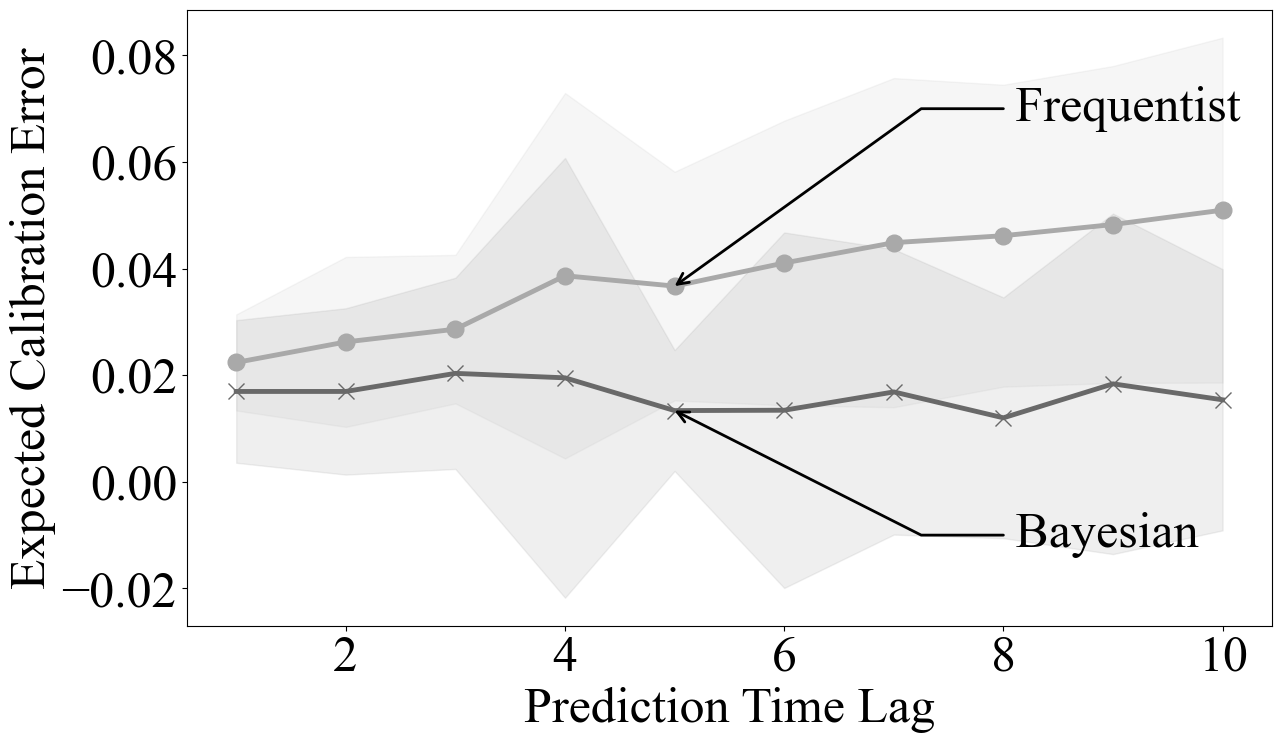}
            \vspace{-0.6cm}
            \caption{}
            \label{fig:prediction_ece}
        \end{subfigure}
        \hfill
        \begin{subfigure}{\textwidth}
            \centering
            \includegraphics[width=\textwidth]{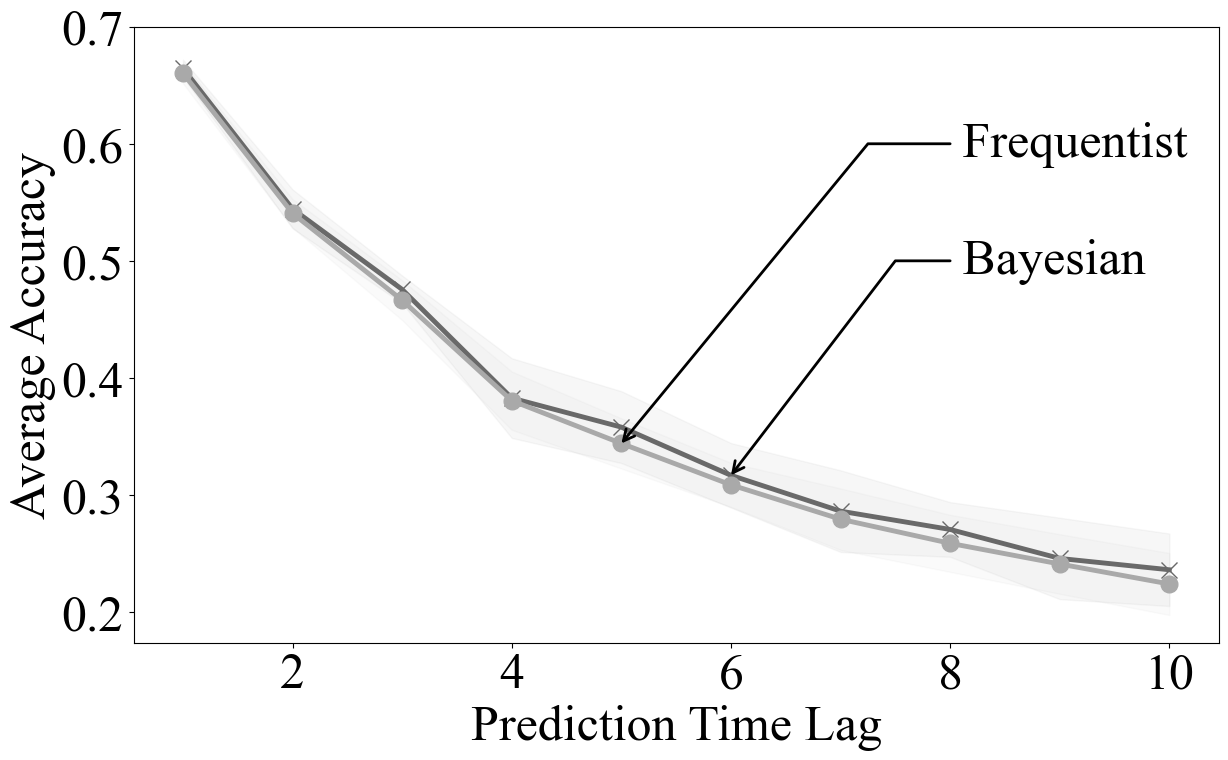}
            \caption{}
            \label{fig:prediction_accuracy}
        \end{subfigure}
    \end{subfigure}
    \caption{
    Reliability plots for packet drop prediction with time lag $T^{\mathrm{H}} = 4$ for (a) the frequentist MAP model and (b) the Bayesian model.
    (c) Expected calibration error and (d) accuracy of the predictions for both Bayesian and frequentist models as a function on the prediction time lag $T^{\mathrm{H}}$, ranging from $T^{\mathrm{H}} = 1$ to $T^{\mathrm{H}} = 10$.
    All the results are averaged over $20$ independently learned models.
    }
    \label{fig:prediction_metrics}
\end{figure*}

\subsection{Policy Evaluation}\label{subsec:experimental_results_policy_evaluation}

In this section, we evaluate the performance of policy optimization in the ground-truth environment by using the following metrics:
\textit{(i)} the \emph{throughput}, i.e., the average number of packets successfully sent at each time step  (Fig. \ref{fig:throughput_per_model_steps}); 
and \textit{(ii)} the average probability of \emph{buffer overflow} across all devices (Fig. \ref{fig:overflow_per_model_steps}).
We focus on the impact of the size of the model learning dataset $\mathcal{D}^{\pi_d}_T$ by varying the number of random data collection steps $T$ from $0$ to $20$ prior to the model learning phase.
The results are averaged over $50$ independent data collection, model learning and policy optimization cycles (phases \textcircled{\raisebox{-0.9pt}{1}}, \textcircled{\raisebox{-0.9pt}{2}} and \textcircled{\raisebox{-0.9pt}{3}} in Fig. \ref{fig:DT_workflow_MAC}).

From Fig. \ref{fig:policies_metrics}, we observe that, in regimes with high data availability during the model learning phase, i.e., with large $T$, both Bayesian and frequentist model-based methods yield policies with similar performance to the oracle-aided benchmark.
In the low-data regime, however, Bayesian learning achieves superior performance as compared to its frequentist counterpart with, for instance, a $20\%$ increase in throughput at $T=10$.
With frequentist learning, which disregards epistemic uncertainty, policy optimization is prone to \emph{model exploitation}, whereby the optimized policy is misled by model errors into taking actions that are unlikely to be advantageous in the ground-truth dynamics.
By using an ensemble of models with distinct transition dynamics in state-action space regions with high epistemic uncertainty, Bayesian learning reduces the sensitivity of the optimized policy to model errors.

\subsection{Anomaly Detection}

We now consider the performance of anomaly detection, as defined in Sec. \ref{subsec:anomaly_detection}, by assuming that an anomalous event occurs when device $2$ is disconnected, resulting in an anomalous packet-generation distribution $\Tilde{P}(g_{t+1})$ for which a packet is generated at device $1$ only with probability $0.4$, and no packet is generated either at device $1$ or $2$ with probability $0.6$.
To focus on such anomalies at the packet generation level, we use the log-likelihood $LL( \mathcal{D}^{\pi}_{T^{\mathrm{M}}} | \theta ) = \sum_{t = 1}^{T^\mathrm{M}} \log( T_{\theta^{\mathrm{G}, 1}}(g^{\mathcal{C}^1}_{t+1}))$ in the disagreement metric (\ref{eq:monitoring_metric}).
Furthermore, as mentioned in Sec. \ref{subsec:experimental_benchmarks}, we consider as benchmark a standard test based on the log-likelihoods $LL( \mathcal{D}^{\pi}_{T^{\mathrm{M}}} | \theta^{\mathrm{MAP}} )$ obtained from MAP-based frequentist learning.

For each model learning dataset size $T=20$ and $T=50$, we compute the Bayesian disagreement metrics and frequentist log-likelihoods for $16000$ independently sampled monitoring datasets $\mathcal{D}^{\pi}_{T^{\mathrm{M}}}$ with $T^\mathrm{M} = 1$, where half of the datasets are sampled from the ground-truth distribution under normal circumstances, while the other half is sampled with device $2$ disconnected.
We then report the false positive rates (FPR) and the true positive rates (TPR) of the anomaly detection tests in Fig. \ref{fig:anomaly_detection} by varying the detection threshold.
The experiment is repeated $50$ times over independent data collection and model learning phases (phases \textcircled{\raisebox{-0.9pt}{1}} and \textcircled{\raisebox{-0.9pt}{2}} in Fig. \ref{fig:DT_workflow_MAC}), while the optimized policy $\pi$ used to report experiences $\mathcal{D}^{\pi}_{T^{\mathrm{M}}}$ remains the same.

For both model-learning dataset sizes of $T = 20$ and $T = 50$ steps in Fig. \ref{fig:auc_anomaly_detection}, Bayesian anomaly detection achieves, on average, a higher area under the receiver operating characteristic (ROC) curve; with a $5\%$ average area increase and a $22\%$ larger area at the $25\%$ quantile for $T=20$ compared to its frequentist counterpart.
From Fig. \ref{fig:roc_anomaly_detection}, the proposed Bayesian framework is also observed to uniformly outperform the frequentist ROC curve for $T = 20$ steps, while providing higher performance at lower FPR for $T = 50$ steps.
For instance, at a TPR of $0.75$ in Fig. \ref{fig:roc_anomaly_detection}, the Bayesian anomaly detector has a FPR of $0.30$ for a model learning dataset size of $T=20$ and a FPR of $0.15$ for a dataset size of $T=50$; whereas the frequentist benchmark has a FPR of $0.34$ for $T=20$ and $0.21$ for $T=50$.
These results suggest that measuring epistemic uncertainty, instead of likelihood, can yield more effective and robust monitoring solutions.

\subsection{Prediction} \label{subsec:experimental_results_prediction}

\begin{figure}[t]
    \centering
    \begin{subfigure}{0.49\textwidth}
        \centering
        \includegraphics[width=\textwidth]{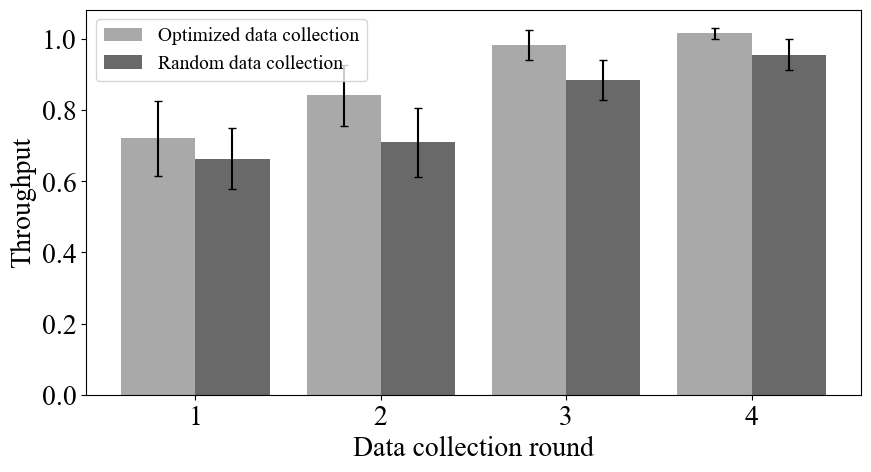}
        \vspace{-0.6cm}
        \caption{}
        \label{fig:directed_exploration_throughput}
    \end{subfigure}
    \hfill
    \begin{subfigure}{0.49\textwidth}
        \centering
        \vspace{+0.2cm}
        \includegraphics[width=\textwidth]{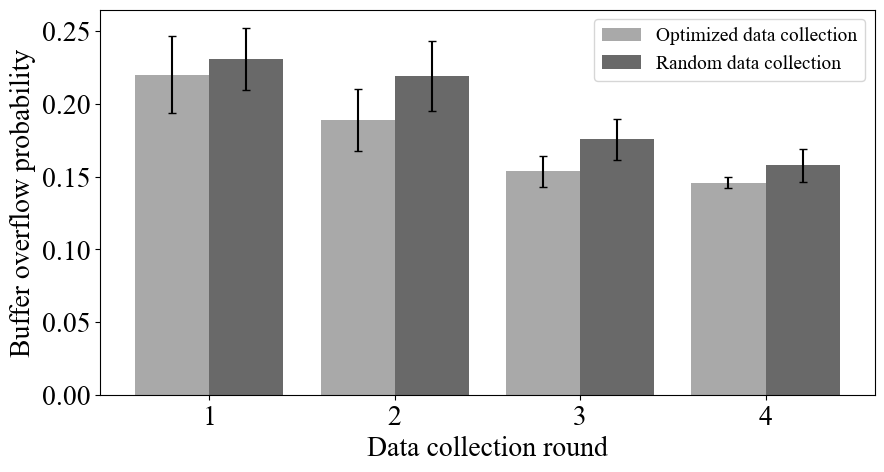}
        \vspace{-0.6cm}
        \caption{}
        \label{fig:directed_exploration_overflow}
    \end{subfigure}
    \caption{
    (a) Throughput and (b) buffer overflow probability as a function of the number of data collection rounds using a random (dark gray) and an optimized (light gray) data collection policy, as described in Sec. \ref{subsec:data_collection_optimization}.
    All the results are averaged over $50$ independent data collection, model learning and policy optimization cycles.
    }
    \label{fig:directed_exploration}
\end{figure}

In this section, we are interested in predicting the number of packet drops, i.e., buffer overflows, experienced across all devices starting from a uniformly sampled state $s_1$.
We collect $T= 100$ data samples using a random data collection policy $\pi_d$, train a Bayesian model $P(\theta | \mathcal{D}^{\pi_d}_{T})$, and use it to produce an optimized policy $\pi$ as described in Sec. \ref{subsec:control_policy_optimization}.
Following Sec. \ref{subsec:prediction}, we define our target metric over the time lag $T^{\mathrm{H}} \in \{1, \dots, 10\}$ as
\begin{equation}
\label{eq:prediction_statistic_case_study}
    y_p =  \sum_{t=1}^{T^{\mathrm{H}}} \sum_{k \in \mathcal{K}} \mathbbm{1}_{\{ q^k_t = Q^k_{\mathrm{max}}, g^k_{t+1} = 1, d^k_{t+1} = 0 \}},
\end{equation}
where the state variables of future trajectories in $\mathcal{D}^{\pi}_{T^{\mathrm{H}}}$ are taken with respect to the optimized policy $\pi$.
Note that, since the optimized policy $\pi$ differs from the data collection policy $\pi_d$, the datasets $\mathcal{D}^{\pi_d}_{T}$ and $\mathcal{D}^{\pi}_{T^{\mathrm{H}}}$ are drawn from two distinct distributions.
Therefore, the number of packet drops $y_p$ cannot be predicted from the currently available data $\mathcal{D}^{\pi_d}_{T}$, and the accuracy of the prediction depends on how well the learned model $T_{\theta}$ at the DT can generalize to new, unseen, conditions.

In order to estimate the packet-drop rate $y_p$, we roll out $10$-steps trajectories from $s_1$ using the learned model.
Furthermore, for the Bayesian model, we average the confidence of each prediction over $20$ sampled models $T_{\theta}$ with $\theta \sim P(\theta | \mathcal{D}^{\pi_d}_{T})$, with $100$ trajectories per model; while, for the frequentist MAP benchmark, we only average $100$ trajectories over the single model $T_{\theta^{\mathrm{MAP}}}$ with $\theta^{\mathrm{MAP}} = \argmax_{\theta} P(\theta | \mathcal{D}^{\pi_d}_{T})$.
The predicted outcome $y_p$ in (\ref{eq:prediction_statistic_case_study}) is tested against 100 outcomes sampled from the ground-truth environment with policy $\pi$ and starting state $s_1$.
We average our results over $20$ independent data collection and model learning cycles (phases \textcircled{\raisebox{-0.9pt}{1}} and \textcircled{\raisebox{-0.9pt}{2}} in Fig. \ref{fig:DT_workflow_MAC}) for $200$ uniformly sampled starting states $s_1$.

We evaluate the performance both in terms of accuracy (Fig. \ref{fig:prediction_accuracy}) and calibration (Fig. \ref{fig:prediction_metrics}a-c).
Calibration performance is evaluated using the standard \emph{reliability plot} and \emph{expected calibration error} (ECE) \cite{guo2017calibration}.
As seen in Fig. \ref{fig:prediction_accuracy}, the prediction accuracy of the Bayesian and frequentist approaches are very similar for all values of $T^{\mathrm{H}} \in \{1, \dots, 10\}$.
However, as we increase the prediction time lag $T^{\mathrm{H}}$, the frequentist approach tends to make incorrect decisions with a high level of confidence, while Bayesian learning correctly evaluates its confidence level.  

To see this, we first observe the reliability plots in Fig. \ref{fig:prediction_metrics}a-b, which are obtained for $T^{\mathrm{H}} = 4$. Reliability plots evaluate prediction accuracy as a function of the confidence level of the decision output by the model.
Perfect calibration is obtained when the confidence (light gray) and accuracy (dark gray) bars are equal.
As anticipated, the frequentist model is observed to be overconfident, while the Bayesian model provides a good match between confidence and accuracy at all confidence levels with a meaningful rate of occurrence (displayed at the bottom of the reliability plots).
The ECE, which evaluates the average difference between confidence and accuracy \cite{guo2017calibration} (Fig. \ref{fig:prediction_ece}), confirms the advantages of Bayesian learning in terms of quality of uncertainty quantification. 

\subsection{Data Collection Optimization}

An optimized data collection policy, as described in Sec. \ref{subsec:data_collection_optimization}, can be useful to improve the estimate of the channel distribution $P(d_{t+1} | a_t)$ since the latter can be explored by controlling the number of transmitted packets.
In this last experiment, we evaluate the advantages of data collection policy optimization across four data collection rounds.

During each round $i \in \{1,2,3,4\}$, the DT collects information about $T^d = 5$ transitions in the ground-truth environment using the data collection policy $\pi_{d, i}$.
The latter is optimized as discussed in Sec. \ref{subsec:data_collection_optimization} using the available data $\mathcal{D}_{ \leq i-1} = \bigcup_{j=1}^{i-1} \mathcal{D}^{\pi_{d, j}}_{T^d}$.
Note that in this problem the data collection reward (\ref{eq:exploration_reward_max}) can be evaluated in closed form using the digamma function \cite{scutari2018dirichlet}.

We evaluate the advantage of the optimized data collection scheme by training a control policy $\pi$ as detailed in Sec. \ref{subsec:control_policy_optimization} using the model $P(\theta | \mathcal{D}_{\leq i})$ available at the end of each round, and evaluating its performance in the ground-truth environment in terms of throughput (Fig. \ref{fig:directed_exploration_throughput}) and buffer overflow probability (Fig. \ref{fig:directed_exploration_overflow}), as described in Sec. \ref{subsec:experimental_results_policy_evaluation}.
The results presented in Fig. \ref{fig:directed_exploration} are averaged over 50 independent data collection (with and without optimization), model learning and policy optimization cycles (phases \textcircled{\raisebox{-0.9pt}{1}}, \textcircled{\raisebox{-0.9pt}{2}}, \textcircled{\raisebox{-0.9pt}{3}} and \textcircled{\raisebox{-0.9pt}{4}} in Fig. \ref{fig:DT_workflow_MAC}).

Since the data collection policy $\pi_{d, 1}$ is trained using the prior model $P(\theta)$ during the first round, the models $T_{\theta}$ with $\theta \sim P(\theta)$ tend to disagree under most transitions, and the performance of the optimized data collection scheme is close to its random counterpart.
However, after the first round, the data collection reward (\ref{eq:exploration_reward_max}) is able to target a smaller subset of transitions with higher epistemic uncertainty, yielding a $18.5\%$ increase in throughput at the end of the second round compared to random exploration.
As the number of rounds increases, the performance gap between the two collection strategies is reduced and we approach the optimal performance of the oracle-aided benchmark in Fig. \ref{fig:policies_metrics}.

\section{Conclusions} \label{sec:conclusion}

This paper has proposed a Bayesian framework for the development of a DT platform aimed at the control, monitoring, and analysis of a communication system.
By accounting for model uncertainty via ensembling, and compared to conventional single-model approaches, the proposed Bayesian DT framework was shown to obtain more reliable performance for multi-agent RL-based control, prediction, anomaly detection, and data collection in the regime of limited data available at the DT from the PT.
For some quantitative examples, we demonstrated a $20\%$ increase in throughput for multi-access transmission from IoT devices, with an additional $18.5\%$ increase obtained by using an optimized data-collection policy; a $5\%$ larger area under the ROC curve for anomaly detection; and a reduction by half of the calibration error for prediction.
Future work may investigate the application of the Bayesian DT framework to other use cases in telecommunication \cite{sheen2020digital, dai2020deep}; the use of more complex policies accounting for partial observability at each agent \cite{zhu2017improving}; as well as the presence of multiple interacting DTs and/or PTs \cite{wu2021digital}, along with the optimal allocation of DTs across cloud and edge \cite{lu2021adaptive, jagannath2022digital}.

\appendices

\section{Table of Notations} \label{apx:table_notations}

\begin{table}[H]
\centering
\renewcommand{\arraystretch}{1.2}
\begin{tabular}{ | m{0.13\textwidth} | m{0.30\textwidth} | }

    \hline
    \textbf{Notation} & \textbf{Meaning} \\
    \hline
    $K$ & Number of agents in the PT system \\
    \hline
    $o^k_t$ & Observation of agent $k$ at time step $t$ \\
    \hline
    $s_t$ & Overall state of the PT system at time step $t$ \\
    \hline
    $s^i_t$ & $i\mbox{-th}$ subset of state variables of the PT system at time step $t$ \\
    \hline
    $a^k_t$ & Action of agent $k$ at time step $t$ \\
    \hline
    $a_t$ & Joint action of all agents at time step $t$ \\
    \hline
    $h^k_t$ & Action-observation history of agent $k$ up to time step $t$ \\
    \hline
    $\pi^k(a^k_t | h^k_t)$ & Policy of agent $k$ \\
    \hline
    $\pi(a_t | s_t)$ & Decentralized policy of all agents \\
    \hline
    $\pi^k_d(a^k_t | h^k_t)$ & Data-collection policy of agent $k$ \\
    \hline
    $\pi_d(a_t | s_t)$ & Data-collection policy of all agents \\
    \hline
    $T(s_{t+1} | s_t, a_t)$ & PT ground-truth transition probability  \\
    \hline
    $T_{\theta}(s_{t+1} | s_t, a_t)$ & DT model of the PT transition probabilities with parameter $\theta$  \\
    \hline
    $T^i_{\theta^i} (s^i_{t+1} | s^{>i}_t, a^{>i}_t)$ & DT model of the transition probabilities of the $i\mbox{-th}$ state subset with parameter $\theta^i$ \\
    \hline
    $\mathcal{D}^{\pi_d}_{T}$ & Dataset containing $T$ transitions collected from the PT system under policy $\pi_d$ \\
    \hline
    $r(s_t, a_t, s_{t+1})$ & Reward function for transition $(s_t, a_t, s_{t+1})$ \\
    \hline
    $r_e(s_t, a_t, s_{t+1})$ & Reward function with exploration bonus \\
    \hline
    $\alpha_e$ & Exploration bonus temperature parameter \\
    \hline
    $r_d(s_t, a_t)$ & Data collection reward function \\
    \hline
    $G_t$ & Total discounted return from time step $t$ \\
    \hline
    $\gamma$ & Discounting factor \\
    \hline
    $T^{\mathrm{M}}$ & Monitoring time window \\
    \hline
    $\mathcal{D}^{\pi}_{T^{\mathrm{M}}}$ & Monitoring dataset containing $T^{\mathrm{M}}$ transitions collected from the PT under policy $\pi$ \\
    \hline
    $T^{\mathrm{H}}$ & Prediction time lag \\
    \hline
    $\mathcal{D}^{\pi}_{T^{\mathrm{H}}}$ & Predicted PT trajectory under policy $\pi$ containing $T^{\mathrm{H}}$ transitions \\
    \hline
    $y_p$ & Prediction target metric \\
    \hline
\end{tabular}
\caption{Notations}
\label{tab:notations}
\end{table}


\ifCLASSOPTIONcaptionsoff
  \newpage
\fi

\bibliographystyle{IEEEtran}
\bibliography{refs}


\begin{IEEEbiography}[{\includegraphics[width=1in,height=1.25in,clip,keepaspectratio]{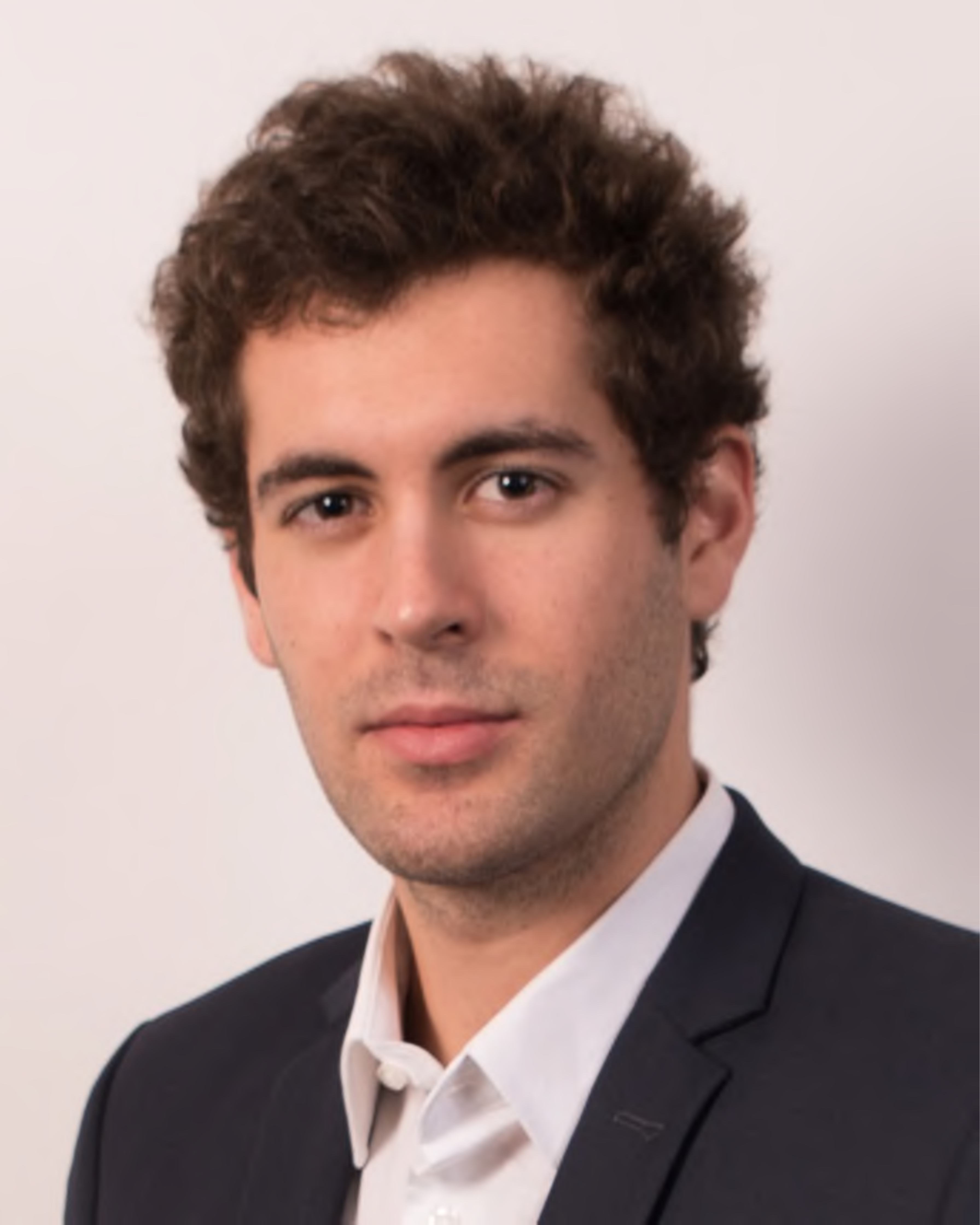}}]{Clement Ruah}
is currently pursuing his Ph.D. degree in Machine Learning at King's College London, United Kingdom.
He received his postgraduate degree in Engineering from the French ``grande ecole'' CentraleSupelec, and a Master's degree in Biomedical Engineering (neuroscience stream) with distinction from Imperial College London in 2018.
His research interest include model-based machine learning, Bayesian learning and multi-agent reinforcement learning.
\end{IEEEbiography}

\begin{IEEEbiography}[{\includegraphics[width=1in,height=1.25in,clip,keepaspectratio]{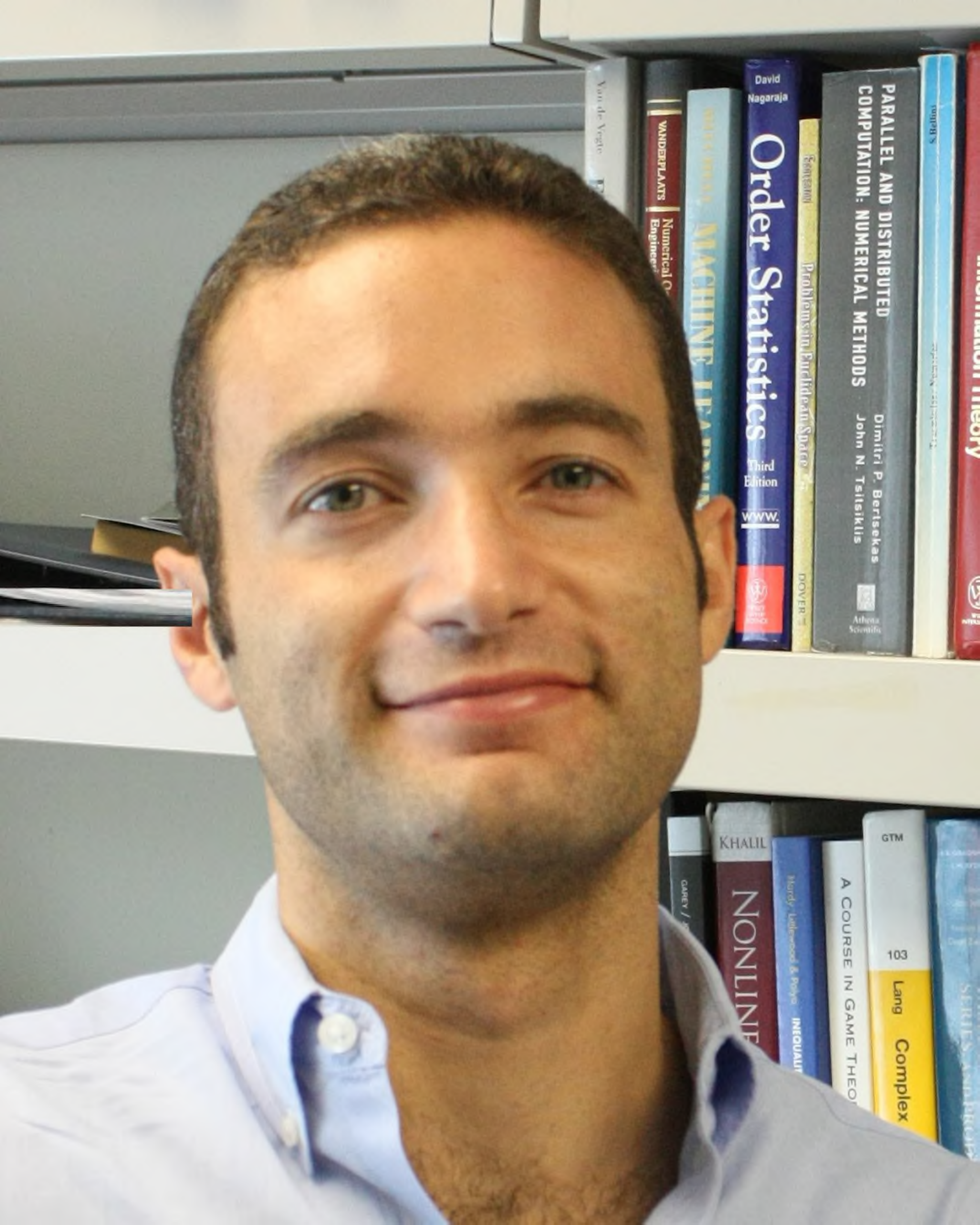}}]{Osvaldo Simeone}
is a Professor of Information Engineering with the Centre for Telecommunications Research at the Department of Engineering of King's College London, where he directs the King's Communications, Learning and Information Processing lab. He received an M.Sc. degree (with honors) and a Ph.D. degree in information engineering from Politecnico di Milano, Milan, Italy, in 2001 and 2005, respectively.
From 2006 to 2017, he was a faculty member of the Electrical and Computer Engineering (ECE) Department at New Jersey Institute of Technology (NJIT), where he was affiliated with the Center for Wireless Information Processing (CWiP). His research interests include information theory, machine learning, wireless communications, neuromorphic computing, and quantum machine learning.
Dr Simeone is a co-recipient of the 2022 IEEE Communications Society Outstanding Paper Award, the 2021 IEEE Vehicular Technology Society Jack Neubauer Memorial Award, the 2019 IEEE Communication Society Best Tutorial Paper Award, the 2018 IEEE Signal Processing Best Paper Award, the 2017 JCN Best Paper Award, the 2015 IEEE Communication Society Best Tutorial Paper Award and of the Best Paper Awards of IEEE SPAWC 2007 and IEEE WRECOM 2007.
He was awarded an Open Fellowship by the EPSRC in 2022 and a Consolidator grant by the European Research Council (ERC) in 2016.
His research has been also supported by the U.S. National Science Foundation, the European Commission, the European Research Council, the Vienna Science and Technology Fund, the European Space Agency, as well as by a number of industrial collaborations including with Intel Labs and InterDigital.
He was the Chair of the Signal Processing for Communications and Networking Technical Committee of the IEEE Signal Processing Society in 2022, as well as of the UK \& Ireland Chapter of the IEEE Information Theory Society from 2017 to 2022.
He was a Distinguished Lecturer of the IEEE Communications Society in 2021 and 2022, and he was a Distinguished Lecturer of the IEEE Information Theory Society in 2017 and 2018.
Prof. Simeone is the author of the textbook "Machine Learning for Engineers" published by Cambridge University Press, four monographs, two edited books, and more than 200 research journal and magazine papers.
He is a Fellow of the IET, EPSRC, and IEEE.
\end{IEEEbiography}

\begin{IEEEbiography}[{\includegraphics[width=1in,height=1.25in,clip,keepaspectratio]{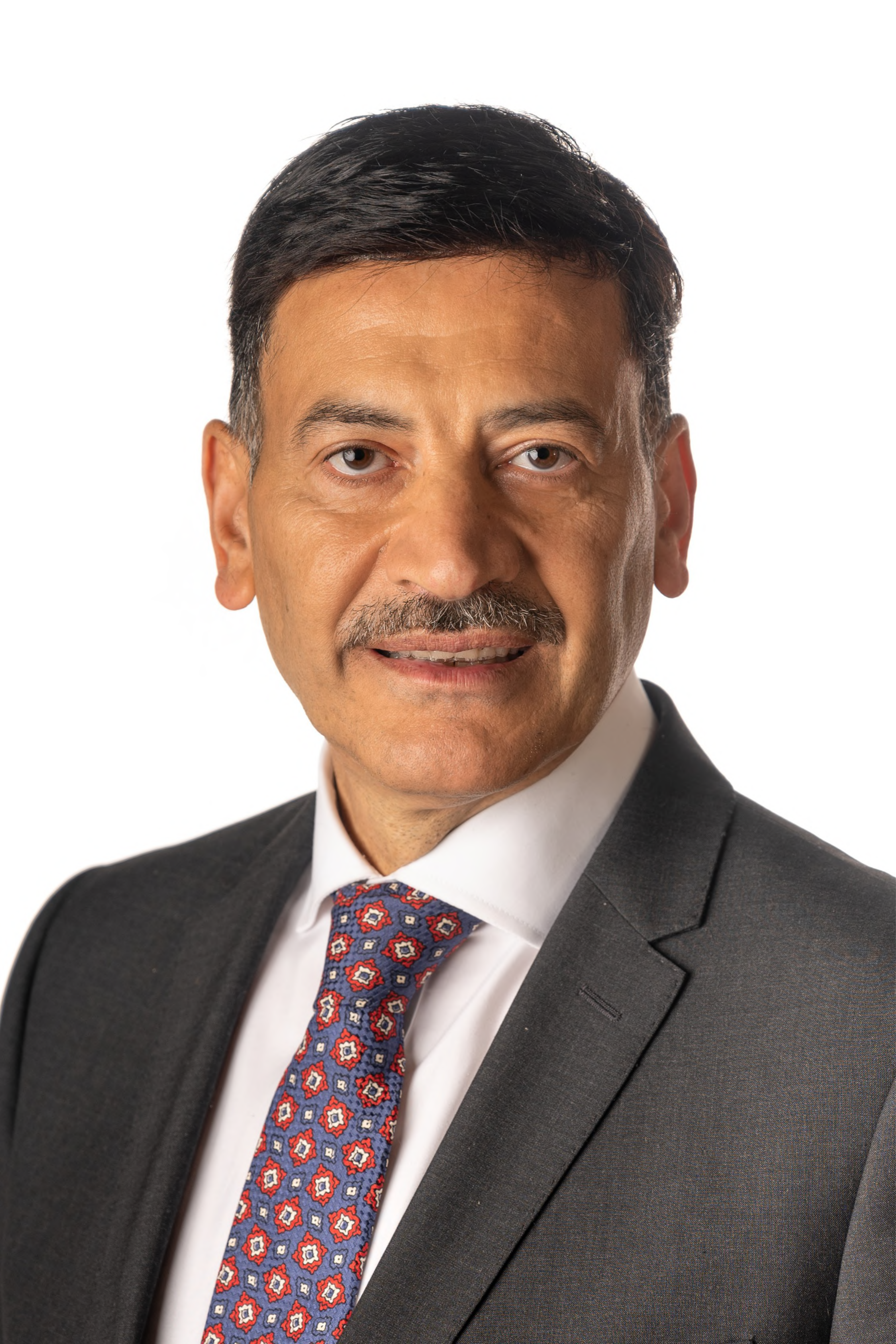}}]{Bashir M. Al-Hashimi}
(Fellow, IEEE) is an ARM Professor of computer engineering and Vice President (Research \& Innovation) of King’s College London.
He worked in the electronics design industry for eight years prior to embarking on an academic career with the School of Electronics and Computer Science, in the University of Southampton, in 1999.
In 2008, he founded the Arm-ECS industry-academia centre of research excellence in energy-efficient computing.
As an interdisciplinary researcher, he has successfully led a number of large-scale interdisciplinary research programmes funded by the EPSRC and industry.
He has supervised 40 Ph.D. students to successful completion, published 380 referred technical papers, and authored or co-authored seven books. 
\end{IEEEbiography}

\end{document}